\documentclass[sigconf,dvipsnames]{acmart}

\usepackage{microtype}
\usepackage{graphicx}
\usepackage{subfigure}
\usepackage{amsmath}
\usepackage{commath}
\usepackage[mathscr]{euscript}
\usepackage{dsfont}

\usepackage{paralist}
\usepackage{geometry}
\usepackage{booktabs} 

\usepackage{ulem}
\usepackage[margin=1cm]{caption}

\usepackage{hyperref}

\newcommand{\Review}[1]{#1}

\usepackage{blindtext}
\usepackage{todonotes}
\usepackage{xfrac}

\usepackage{multirow}
	
\usepackage{color, colortbl}

\author[T. Gaudelet \textit{et al.}]{Thomas Gaudelet$^1$, Ben Day$^{1,2}$, Arian R. Jamasb$^{1,2,3}$, Jyothish Soman$^1$, Cristian Regep$^1$, Gertrude Liu$^1$, Jeremy B. R. Hayter$^1$, Richard Vickers$^1$, Charles Roberts$^{1,4}$, Jian Tang$^{5,6}$, David Roblin$^{1,4,7}$,\\ Tom L. Blundell$^3$, Michael M. Bronstein$^{8,9}$, and Jake P. Taylor-King$^{1,4}$ \\[1ex]
\large{$^1$ Relation Therapeutics, London, UK $\quad$ $^2$ The Computer Laboratory, University of Cambridge, UK \\ \quad $^3$ Department of Biochemistry, University of Cambridge, UK \quad  $^4$ Juvenescence, London, UK  \\ \quad $^5$ Mila, the Quebec AI Institute, Canada \quad $^6$ HEC Montreal, Canada  \quad $^7$ The Francis Crick Institute, London, UK \\ \quad $^8$ Department of Computing, Imperial College London, UK \quad $^9$ Twitter, UK} \\ \texttt{jake@relationrx.com} }

\begin{document}

\title[\bf Utilising Graph Machine Learning within Drug Discovery and Development]{{\bf Utilising Graph Machine Learning within Drug Discovery and Development}}

\begin{abstract}
Graph Machine Learning (GML) is receiving growing interest within the pharmaceutical and biotechnology industries for its ability to model biomolecular structures, the functional relationships between them, and integrate multi-omic datasets --- amongst other data types. Herein, we present a multidisciplinary academic-industrial review of the topic within the context of drug discovery and development. After introducing key terms and modelling approaches, we move chronologically through the drug development pipeline to identify and summarise work incorporating: target identification, design of small molecules and biologics, and drug repurposing. Whilst the field is still emerging, key milestones including repurposed drugs entering \textit{in vivo} studies, suggest graph machine learning will become a modelling framework of choice within biomedical machine learning.
\end{abstract}

\twocolumn[{%
\renewcommand\twocolumn[1][]{#1}%
\maketitle
}]

\section{Introduction}

\begin{figure*}[ht]
    \centering
    \includegraphics[width=0.99\linewidth]{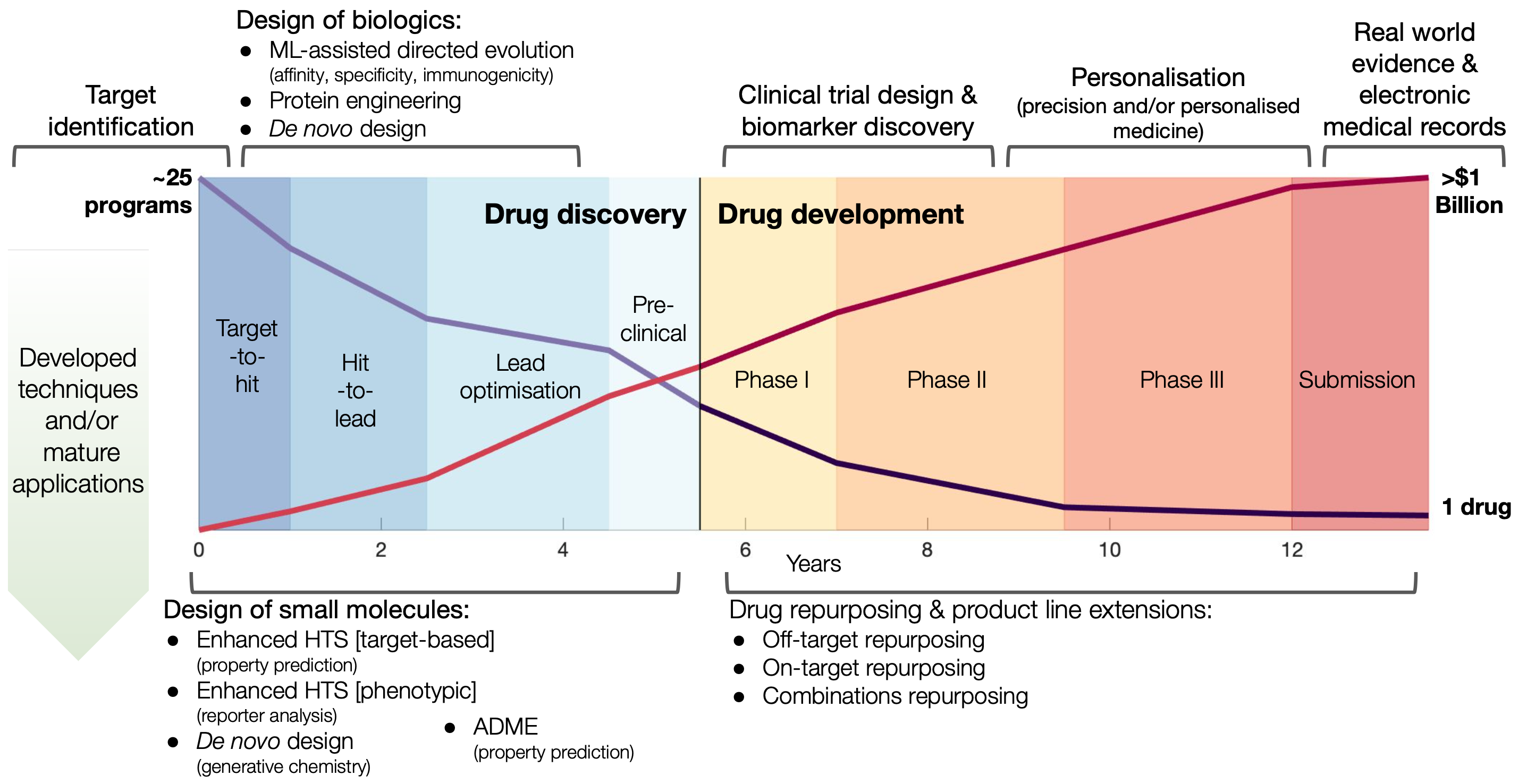}
    \caption{Timeline of drug development linked to potential areas of application by GML methodologies. \Review{Preclinical drug discovery applications are shown on the left side of the figure ($\sim$5.5 years), and clinical drug development applications are shown on the right hand side of the figure ($\sim$8 years). Over this period, for every $\sim$25 drug discovery programs, a single successful drug reaches market approval. Applications listed in the top of half of the figure are less developed in the context of GML with limited experimental validation.} Financial, timeline and success probability data taken from Paul \textit{et al.} \cite{paul2010improve}.}
    \label{fig:drugdev}
\end{figure*}

The process from drug discovery to market costs, on average, well over \$1 billion and can span 12 years or more \cite{dimasi2016innovation,Deloitte2019,wouters2020estimated}; due to high attrition rates, rarely can one progress to market in less than ten years \cite{martin2017clinical,paul2010improve}. The high levels of attrition throughout the process not only make investments uncertain but require market approved drugs to pay for the earlier failures. Despite an industry-wide focus on efficiency for over a decade, spurred on by publications and annual reports highlighting revenue cliffs from ending exclusivity and falling productivity, significant improvements have proved elusive against the backdrop of scientific, technological and regulatory change \cite{Deloitte2019}. For the aforementioned reasons, there is now a greater interest in applying computational methodologies to expedite various parts of the drug discovery and development pipeline \cite{reda2020machine}, see Figure \ref{fig:drugdev}.

Digital technologies have transformed the drug development process generating enormous volumes of data. Changes range from moving to electronic lab notebooks \cite{nishida2020description}, electronic regulatory submissions, through increasing volumes of laboratory, experimental, and clinical trial data collection \cite{SuraeDDW} including the use of devices \cite{coran2019advancing,marquis2019technology} to precision medicine and the use of ``big data'' \cite{hulsen2019big}. The data collected about therapies extends well beyond research and development to include hospital, specialist and primary care medical professionals' patient records --- including observations taken from social media, e.g. for pharmacovigilance \cite{sloane2015social,sarker2015utilizing}. There are innumerable online databases and other sources of information including scientific literature, clinical trials information, through to databases of repurposable drugs \cite{Corsello2017, pantziarka2018redo_db}. Technological advances now allow for greater -omic profiling beyond genotyping and whole genome sequencing (WGS); standardisation of microfluidics and antibody tagging has made single-cell technologies widely available to study both the transcriptome, e.g. using RNA-seq \cite{heath2016single}, the proteome (targeted), e.g. via mass cytometry \cite{spitzer2016mass}, or even multiple modalities together \cite{mcginnis2019multi}.

One of the key characteristics of biomedical data that is produced and used in the drug discovery process is its inter-connected nature. Such data structure can be represented as a graph, a mathematical abstraction ubiquitously used across disciplines and fields in biology to model the diverse interactions between biological entities that intervene at the different scales. At the molecular scale, proteins and other biomolecules can be represented as graphs capturing spatial and structural relationships between their amino acid residues \cite{fout2017protein,zamora2019structural} and small molecule drugs as graphs relating their constituent atoms and chemical bonding structure \cite{duvenaud2015convolutional,klicpera2020directional}. At an intermediary scale, interactomes are graphs that capture specific types of interactions between biomolecular species (e.g. metabolites, mRNA, proteins) \cite{han2008understanding}, with protein--protein interaction (PPI) graphs being perhaps most commonplace. Finally, at a higher level of abstraction, knowledge graphs can represent the complex relationships between drugs, side effects, diagnosis, associated treatments, and test results \cite{zhu2019graph, choi2020learning} as found in electronic medical records (EMR).

Within the last decade, two emerging trends have reshaped the data modelling community: network analysis and deep learning. The ``network medicine'' paradigm has long been recognised in the biomedical field \cite{barabasi2011network}, with multiple approaches borrowed from graph theory and complex network science applied to biological graphs such as PPIs and gene regulatory networks (GRNs). Most approaches in this field were limited to \textit{handcrafted} graph features such as centrality measures and clustering. In contrast, deep neural networks, a particular type of machine learning algorithms, are used to learn optimal tasks-specific features. The impact of deep learning was ground-breaking in computer vision \cite{voulodimos2018deep} and natural language processing \cite{young2018recent} but was limited to specific domains by the requirements on the regularity of data structures. At the convergence of these two fields is Graph Machine Learning (GML) a new class of ML methods exploiting the structure of graphs and other irregular datasets (point clouds, meshes, manifolds, etc).

The essential idea of GML methods is to learn effective feature representations of nodes~\cite{perozzi2014deepwalk, sun2018rotate} (e.g., users in social networks), edges (e.g. predicting future interactions in recommender systems), or the entire graphs~\cite{sun2019infograph} (e.g. predicting properties of molecular graphs). In particular, a type of methods called graph neural networks (GNNs)~\cite{kipf2016semi, velivckovic2018graph, gilmer2017neural}, which are deep neural network architectures specifically designed for graph-structure data, are attracting growing interest. GNNs iteratively update the features of the nodes of a graph by propagating the information from their neighbours. These methods have already been successfully applied to a variety of tasks and domains such as recommendation in social media and E-commerce \cite{pal2020pinnersage, yang2019aligraph,rossi2020sign,rossi2020temporal}, traffic estimations in Google Maps \cite{google2020}, misinformation detection in social media \cite{monti2019fake}, and various domains of natural sciences including modelling of glass \cite{cranmer2020discovering} and event classification in particle physics \cite{shlomi2020graph,choma2018graph}. 


In the biomedical domain, GML has now set the state of the art for mining graph-structured data including drug--target--indication interaction and relationship prediction through knowledge graph embedding \cite{sun2018rotate,schlichtkrull2018modeling,balazevic2019tucker}; molecular property prediction \cite{duvenaud2015convolutional,klicpera2020directional}, including the prediction of absorption, distribution, metabolism, and excretion (ADME) profiles \cite{feinberg2020improvement}; early work in target identification \cite{pittala2020relation} to \textit{de novo} molecule design \cite{zhavoronkov2019deep,gainza2020deciphering}. Most notably, Stokes \textit{et al.} \cite{stokes2020deep} used directed message passing graph neural networks operating on molecular structures to propose repurposing candidates for antibiotic development, validating their predictions \textit{in vivo} to propose suitable repurposing candidates remarkably structurally distinct from known antibiotics. Therefore, GML methods appear to be extremely promising in applications across the drug development pipeline. 

The rest of this paper is organised as follows. Section \ref{notations} introduces notation and mathematical preliminaries to simplify the exposition. Section \ref{mlg} introduces the diverse tasks addressed with GML approaches as well as the main classes of existing methods. Section \ref{sec:drugdev} focuses on the application of machine learning on graphs within drug discovery and development before providing a closing discussion in Section \ref{sec:discussion}. \Review{Compared to previous general review papers on GML \cite{nickel2015review, zhou2018graph, wu2020comprehensive, hamilton2017representation, zhang2020deep} for the machine learning community, the focus of our paper is on biomedical researchers without extensive ML backgrounds as well as ML experts interested in biomedical applications. We provide an introduction to the key terms and  building blocks of graph learning architectures (Sections \ref{notations} \& \ref{mlg}) and contextualise these methodologies within the drug discovery and development pipeline from an industrial perspective for method developers without extensive biological expertise (Section \ref{sec:drugdev}).}

\section{Definitions}\label{notations}

\subsection{Notations and preliminaries of graph theory}

We denote a graph $G=(\mathcal{V},\mathcal{E},\mathbf{X}^v,\mathbf{X}^e)$ where $\mathcal{V}$ is a set of $n = |\mathcal{V}|$ nodes, or vertices, and $\mathcal{E}\subseteq \mathcal{V}\times \mathcal{V}$ is a set of $m$ edges. Let $v_i \in \mathcal{V}$ denote a node and $e_{ij} = (v_i,v_j) \in \mathcal{E}$ denote an edge from node $v_i$ to node $v_j$. When multiple edges can connect the same pair of nodes, the graph is called a \textit{multigraph}. Node features are represented by $\mathbf{X}^v \in \mathbb{R}^{n \times d}$ and $\mathbf{x}^v_i \in \mathbb{R}^{d}$ are the $d$ features of node $v_i$. Edge features, or attributes, are similarly represented by $\mathbf{X}^e\in \mathbb{R}^{m \times c}$ where $\mathbf{x}^e_{i,j} = \mathbf{x}^e_{v_i,v_j} \in \mathbb{R}^c$. We may also denote different nodes as $u$ and $v$ such that $e_{u,v}=(u,v)$ is the edge from $u$ to $v$ with attributes $\mathbf{x}^e_{u,v}$.
Note that under this definition, undirected graphs are defined as directed graphs with each undirected edge represented by two directed edges.

The neighbourhood $\mathcal{N}(v)$ of node $v$, sometimes referred to as \textit{one-hop neighbourhood}, is the set of nodes that are connected to it by an edge, $\mathcal{N}(v) = \{u \in \mathcal{V}|(v,u) \in \mathcal{E}\}$, with shorthand $\mathcal{N}(v_i)=\mathcal{N}_i$ used for compactness. The cardinality of a node's neighbourhood is called its degree and the diagonal degree matrix, $\mathbf{D}$, has elements $\mathbf{D}_{ii}=|\mathcal{N}_i|$.

Two nodes $v_i$ and $v_j$ in a graph $G$ are \textit{connected} if there exists a \textit{path} in $G$ starting at one and ending at the other, i.e. there exists a sequence of consecutive edges of $G$ connecting the two nodes. A graph is connected if there exists a path between every pair of nodes in the graph. The shortest path distance between $v_i$ and $v_j$ is defined as the number of edges in the shortest path between the two nodes and denoted by $d(v_i,v_j)$.

A graph $S=(\tilde{\mathcal{V}},\tilde{\mathcal{E}},\mathbf{X}^{\Tilde{v}},\mathbf{X}^{\Tilde{e}})$ is a \textit{subgraph} of $G$ if and only if $\Tilde{\mathcal{V}}\subseteq \mathcal{V}$ and $\Tilde{\mathcal{E}}\subseteq \mathcal{E}$. If it also holds that $\Tilde{\mathcal{E}} = \left(\Tilde{\mathcal{V}}\times\Tilde{\mathcal{V}}\right)\cap \mathcal{E}$, then $S$ is called an \textit{induced subgraph} of $G$.

The adjacency matrix $\mathbf{A}$ typically represents the relations between nodes such that the entry on the $i^{th}$ row and $j^{th}$ column indicates whether there is an edge from node $i$ to node $j$, with $1$ representing that there is an edge, and $0$ that there is not (i.e. $\mathbf{A}_{ij}=\mathds{1}(v_i,v_j)$). Most commonly, the adjacency matrix is a square (from a set of nodes to itself), but the concept extends to bipartite graphs where an $N \times M$ matrix can represent the edges from one set of $N$ nodes to another set of size $M$, and is sometimes used to store scalar edge weights. The Laplacian matrix of a simple (unweighted) graph is $\mathbf{L} = \mathbf{D} - \mathbf{A}$. The normalised Laplacian $\mathbf{\mathcal{L}} = \mathbf{I} - \mathbf{D}^{\sfrac{-1}{2}}\mathbf{A}\mathbf{D}^{\sfrac{-1}{2}}$ is often preferred, with a variant defined as $\Tilde{\mathbf{\mathcal{L}}} = \mathbf{I} - \mathbf{D}^{-1}\mathbf{A}$.

\subsection{Knowledge graph}

The term \textit{knowledge graph} is used to qualify a graph that captures $r$ types of relationships between a set of entities. In this case, $\mathbf{X}^e$ includes relationship types as edge features. Knowledge graphs are commonly introduced as sets of triplets $(v_i,k,v_j)\in \mathcal{V}\times\mathcal{R}\times \mathcal{V}$, where $\mathcal{R}$ represents the set of relationships. Note that multiple edges of different types can connect two given nodes. As such, the standard adjacency matrix is ill-suited to capture the complexity of a knowledge graph. Instead, a knowledge graph is often represented as a collection of adjacency matrices $\{\mathbf{A}_1,...,\mathbf{A}_r\}$, forming an adjacency tensor, in which each adjacency matrix $\mathbf{A}_i$ captures one type of relationship.

\subsection{Random walks}

A random walk is a sequence of nodes selected at random during an iterative process. A random walk is constructed by considering a random walker that moves through the graph starting from a node $v_i$. At each step, the walker can either move to a neighbouring node with probability $p(v_j|v_i),\-v_j\in\mathcal{N}_i$ or stay on node $v_i$ with probability $p(v_i|v_i)$. The sequence of nodes visited after a fixed number of steps $k$ gives a random walk of length $k$. Graph diffusion is a related notion that models the propagation of a signal on a graph. A classic example is heat diffusion \cite{kondor2002diffusion}, which studies the propagation of heat in a graph starting from some initial distribution.

\subsection{Graph isomorphism}

Two graphs $G=(\mathcal{V}_G,\mathcal{E}_G)$ and $H=(\mathcal{V}_H,\mathcal{E}_H)$ are said to be isomorphic if there exists a bijective function $f: \mathcal{V}_G\mapsto \mathcal{V}_H$ such that $\forall (g_i,g_j) \in \mathcal{E}_G,\- (f(g_i),f(g_j))\in \mathcal{E}_H$. Finding if two graphs are isomorphic is a recurrent problem in graph analysis that has deep ramifications for machine learning on graphs. For instance, in graph classification tasks, it is assumed that a model needs to capture the similarities between pairs of graphs to classify them accurately.

The Weisfeiler-Lehman (WL) graph isomorphism test \cite{wl1968} is a classical polynomial-time algorithm in graph theory. It is based on iterative graph recolouring, starting with all nodes of identical ``colour'' (label). At each step, the algorithm aggregates the colours of nodes and their neighbourhoods and hashes the aggregated colour into unique new colours. The algorithm stops upon reaching a stable colouring. If at that point, the colourings of the two graphs differ, the graphs are deemed non-isomorphic. However, if the colourings are the same, the graphs are possibly (but not necessarily) isomorphic. In other words, the WL test is a necessary but insufficient condition for graph isomorphism. There exist non-isomorphic graphs for which the WL test produces identical colouring and thus considers them \textit{possibly isomorphic}; the test is said to fail in this case \cite{berkholz2015limitations}.

\section{Machine learning on graphs}\label{mlg}

Most machine learning methods that operate on graphs can be decomposed into two parts: a general-purpose encoder and a task-specific decoder \cite{chami2020machine}. The encoder embeds a graph's nodes, or the graph itself, in a low-dimensional feature space. To embed entire graphs, it is common first to embed nodes and then apply a permutation invariant pooling function to produce a graph level representation (e.g. sum, max or mean over node embeddings). The decoder computes an output for the associated task. The components can either be combined in two-step frameworks, with the encoder pre-trained in an unsupervised setting, or in an end-to-end fashion. The end tasks can be classified following multiple dichotomies: supervised/unsupervised, inductive/transductive, and node-level/graph-level.

\paragraph{Supervised/unsupervised task}
This is the classical dichotomy found in machine learning \cite{murphy2012machine}. Supervised tasks aim to learn a mapping function from labelled data such that the function maps each data point to its label and generalises to unseen data points. In contrast, unsupervised tasks highlight unknown patterns and uncover structures in unlabelled datasets. 

\paragraph{Inductive/transductive task} Inductive tasks correspond to supervised learning discussed above. Transductive tasks expect that all data points are available when learning a mapping function, including unlabelled data points \cite{kipf2016semi}. Hence, in the transductive setting, the model learns both from unlabelled and labelled data. In this respect, inductive learning is more general than transductive learning, as it extends to unseen data points.

\paragraph{Node-level/graph-level task} This dichotomy is based on the object of interest. A task can either focus on the nodes within a graph, e.g. classifying nodes within the context set by the graph, or focus on whole graphs, i.e. each data point corresponds to an entire graph \cite{sun2019infograph}. Note that node-level tasks can be further decomposed into node attributes prediction tasks \cite{kipf2016semi} and link inference tasks \cite{sun2018rotate}. The former focuses on predicting properties of nodes while the latter infers missing links in the graph.\\

As an illustration, consider the task of predicting the chemical properties of small molecules based on their chemical structures. This is a graph-level task in a supervised (inductive) setting whereby labelled data is used to learn a mapping from chemical structure to chemical properties. Alternatively, the task of identifying groups of proteins that are tightly associated in a PPI graph is an unsupervised node-level task. However, predicting proteins' biological functions using their interactions in a PPI graph corresponds to a node-level transductive task.

Further, types of tasks can be identified, e.g. based on whether we have static or varying graphs. Biological graphs can vary and evolve along a temporal dimension resulting in changes to composition, structure and attributes \cite{othmer1971instability,praktiknjo2020tracing}. However, the classifications detailed above are the most commonly found in the literature. We review below the existing classes of GML methods.

\subsection{Traditional approaches}\label{s31}

\subsubsection{Graph statistics}

In the past decades, a flourish of heuristics and statistics have been developed to characterise graphs and their nodes. For instance, the diverse centrality measures capture different aspects of graphs connectivity. The \textit{closeness centrality} quantifies how closely a node is connected to all other nodes, and the \textit{betweenness centrality} counts on how many shortest paths between pairs of other nodes in the graph a given node is. Furthermore, graph sub-structures can be used to derive topological descriptors of the wiring patterns around each node in a graph. For instance, motifs \cite{milo2002network}, and graphlets \cite{prvzulj2004modeling} correspond to sets of small graphs used to characterise local wiring patterns of nodes. Specifically, for each node, one can derive a feature vector of length $n$ corresponding to the number of motifs (graphlets) and where the $i^{th}$ entry gives the number of occurrences of motif $i$ in the graph that contains the node.

These handcrafted features can provide node, or graph, representations that can be used as input to machine learning algorithms. A popular approach has been the definition of kernels based on graph statistics that can be used as input to Support Vector Machines (SVM). For instance, the graphlet kernel \cite{shervashidze2009efficient} captures node wiring patterns similarity, and the WL kernel \cite{shervashidze2011weisfeiler} captures graph similarity based on the Weisfeiler-Lehman algorithm discussed in Section \ref{notations}.

\subsubsection{Random walks}

Random-walk based methods have been a popular, and successful, approach to embed a graph's nodes in a low-dimensional space such that node proximities are preserved. The underlying idea is that the distance between node representations in the embedding space should correspond to a measure of distance on the graph, measured here by how often a given node is visited in random walks starting from another node. Deepwalk \cite{perozzi2014deepwalk} and node2vec \cite{grover2016node2vec} are arguably the most famous methods in this category.

In practice, Deepwalk simulates multiple random walks for each node in the graph. Then, given the embedding $\mathbf{x}^v_i$ of a node $v_i$, the objective is to maximise the log probability $\log p(v_j|\mathbf{x}^v_i)$ for all nodes $v_j$ that appear in a random walk within a fixed window of $v_i$. The method draws its inspiration from the SkipGram model developed for natural language processing \cite{mikolov2013efficient}.

DeepWalk uses uniformly random walks, but several follow-up works analyse how to bias these walks to improve the learned representations. For example, node2vec biases the walks to behave more or less like certain search algorithms over the graph. The authors report a higher quality of embeddings with respect to information content when compared to Deepwalk.

\subsection{Geometric approaches}\label{s32}

Geometric models for knowledge graph embedding posit each relation type as a geometric transformation from source to target in the embedding space. Consider a triplet $(s,r,t)$, $s$ denoting the source node and $t$ denoting the target node. A geometric model learns a transformation $\tau(\cdot,\cdot)$ such that $\delta(\tau(\mathbf{h}_s,\mathbf{h}_r),\mathbf{h}_t)$ is small, with $\delta(\cdot,\cdot)$ being some notion of distance (e.g. Euclidean distance) and $\mathbf{h}_x$ denoting the embedding of entity $x$. The key differentiating choice between these approaches is the form of the geometric transformation $\tau$. 

TransE \cite{bordes2013translating} is a purely translational approach, where $\tau$ corresponds to the sum of the source node and relation embeddings. In essence, the model enforces that the motion from the embedding $\mathbf{h}_s$ of the source node in the direction given by the relation embedding $\mathbf{h}_r$ terminates close to the target node's embedding $\mathbf{h}_t$ as quantified by the chosen distance metric. Due to its formulation, TransE is not able to account effectively for symmetric relationships or one-to-many interactions.

Alternatively, RotatE \cite{sun2018rotate} represents relations as rotations in a complex latent space. Thus, $\tau$ applies a rotation matrix $\mathbf{h}_r$, corresponding to the relation, to the embedding vector of the source node $\mathbf{h}_s$ such that the rotated vector $\tau(\mathbf{h}_r,\mathbf{h}_s)$ lies close to the embedding vector $\mathbf{h}_t$ of the target node in terms of Manhattan distance. The authors demonstrate that rotations can correctly capture diverse relation classes, including symmetry/anti-symmetry.

\subsection{Matrix/tensor factorisation}\label{s33}

Matrix factorisation is a common problem in mathematics that aims to approximate a matrix $\mathbf{X}$ by the product of $n$ low-dimensional latent factors, $\mathbf{F}_i,\-i\in\{1,...,n\}$. The general problem can be written as
\begin{align*}
    \mathbf{F}^*_1,...,\mathbf{F}^*_n = \textnormal{argmin}_{\mathbf{F}_1,...,\mathbf{F}_n} \Delta\left(\mathbf{X},\prod_{i=1}^n \mathbf{F}_i\right),
\end{align*}
\noindent where $\Delta(\cdot,\cdot)$ represents a measure of the distance between two inputs, such as Euclidean distance or Kullback-Leibler divergence. 

In machine learning, matrix factorisation has been extensively used for unsupervised applications such as dimensionality reduction, missing data imputation, and clustering. These approaches are especially relevant to the knowledge graph embedding problem and have set state-of-the-art (SOTA) results on standard benchmarks \cite{rossi2020knowledge}. 

For graphs, the objective is to factorise the adjacency matrix $\mathbf{A}$, or a derivative of the adjacency matrix (e.g. Laplacian matrix). It can effectively be seen as finding embeddings for all entities in the graph on a low-dimensional, latent manifold under user-defined constraints (e.g. latent space dimension) such that the adjacency relationships are preserved under dot products.

Laplacian eigenmaps, introduced by Belkin \textit{et al.} \cite{belkin2002laplacian}, is a fundamental approach designed to embed entities based on a similarity derived graph. Laplacian eigenmaps uses the eigendecomposition of the Laplacian matrix of a graph to embed each of the $n$ nodes of a graph $G$ in a low-dimensional latent manifold. The spectral decomposition of the Laplacian is given by equation $\mathbf{L} = \mathbf{Q}\mathbf{\Lambda} \mathbf{Q}^\top$, where $\mathbf{\Lambda}$ is a diagonal matrix with entries corresponding to the eigenvalues of L and column $\mathbf{q}_k$ of $\mathbf{Q}$ gives the eigenvector associated to the $k^{th}$ eigenvalue $\lambda_k$ (i.e. $\mathbf{L} \mathbf{q}_k = \lambda_k \mathbf{q}_k$). Given a user defined dimension $m\leq n$, the embedding of node $v_i$ is given by the vector $(\mathbf{q}_0(i),\mathbf{q}_1(i),\ldots,\mathbf{q}_{m-1}(i))$, where $\mathbf{q}_{*}(i)$ indicates the $i^{th}$ entry of vector $\mathbf{q}_{*}$.

Nickel \textit{et al.} \cite{nickel2011three} introduced RESCAL to address the knowledge graph embedding problem. RESCAL's objective function is defined as \[\mathbf{U}^*,\mathbf{R}^*_i = \textnormal{argmin}_{\mathbf{U},\mathbf{R}_i} \frac{1}{2}\sum_{i=1}^r\|\mathbf{A}_i - \mathbf{U}\mathbf{R}_i\mathbf{U}^\top\|_F^2 + g(\mathbf{U},\mathbf{R}_i),\] where $\mathbf{U}\in\mathbb{R}^{n\times k}$ and $\forall i, \mathbf{R}_i\in\mathbb{R}^{k,k}$, with $k$ denoting the latent space dimension. The function $g(\cdot)$ denotes a regulariser, i.e. a function applying constraints on the free parameters of the model, on the factors $\mathbf{U}$ and $\{\mathbf{R}_1,..,\mathbf{R}_r\}$. Intuitively, factor $\mathbf{U}$ learns the embedding of each entity in the graph and factor $\mathbf{R}_i$ specifies the interactions between entities under relation $i$. Yang \textit{et al.} \cite{yang2014embedding} proposed DistMult, a variation of RESCAL that considers each factor $\mathbf{R}_i$ as a diagonal matrix. Trouillon \textit{et al.} \cite{trouillon2016complex} proposed ComplEx, a method extending DistMult to the complex space taking advantage of the Hermitian product to represent asymmetric relationships.

Alternatively, some existing frameworks leverage both a graph's structural information and the node's semantic information to embed each entity in a way that preserves both sources of information. One such approach is to use a graph's structural information to regularise embeddings derived from the factorisation of the feature matrix $\mathbf{X}^v$ \cite{cai2011graph,chang2015heterogeneous}. The idea is to penalise adjacent entities in the graph to have closer embeddings in the latent space, according to some notion of distance. Another approach is to jointly factorise both data sources, for instance, introducing a kernel defined on the feature matrix \cite{huang2017label,huang2017accelerated}.

\subsection{Graph neural networks}\label{ssec:mlggnns}\label{s34}

Graph Neural Networks (GNNs) were first introduced in the late 1990s \cite{sperduti1997supervised,gori2005new,merkwirth2005automatic} but have attracted considerable attention in recent years, with the number of variants rising steadily \cite{kipf2016semi,hamilton2017inductive,gilmer2017neural,velivckovic2018graph,xu2018powerful,xu2018representation,schlichtkrull2018modeling,maron2018invariant,chami2019hyperbolic}. \Review{From a high-level perspective, GNNs are a realisation of the notion of \textit{group invariance}, a general blueprint underpinning the design of a broad class of deep learning architectures. 
The key structural property of graphs is that the nodes are usually not assumed to be provided in any particular order, and any functions acting on graphs should be \textit{permutation invariant} (order-independent); therefore, for any two isomorphic graphs, the output of said functions are identical. 
A typical GNN consists of one or more layers implementing a node-wise aggregation from the neighbour nodes; since the ordering of the neighbours is arbitrary, the aggregation must be permutation invariant. When applied locally to every node of the graph, the overall function is \textit{permutation equivariant}, i.e. its output is changed in the same way as the input under node permutations. 
}

\Review{GNNs are among the most general class of deep learning architectures currently in existence. Popular architectures such as DeepSets \cite{zaheer2017deep}, transformers \cite{vaswani2017attention}, and convolutional neural networks \cite{lecun2015deep} can be derived as particular cases of GNNs operating on graphs with an empty edge set, a complete graph, and a ring graph, respectively. In the latter case, the graph is fixed and the neighbourhood structure is shared across all nodes; the permutation group can therefore be replaced by the translation group, and the local aggregation expressed as a convolution. 
While a broad variety of GNN architecture exists, their vast majority can be classified into convolutional, attentional, and message-passing ``flavours'' --- with message-passing being the most general formulation. }

\subsubsection{Message passing networks}

\begin{figure}[ht]
    \centering
    \includegraphics[width=0.6\linewidth]{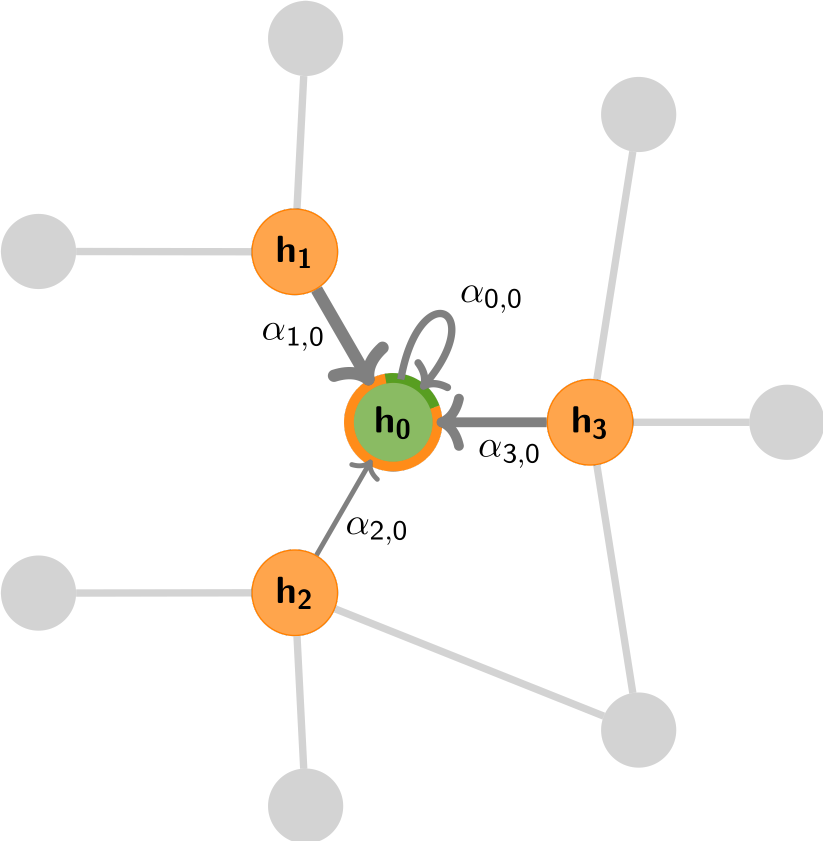}
    \caption{Illustration of a general aggregation step performed by a graph neural network for the central node (green) based on its direct neighbours (orange). Messages may be weighted depending on their content, the source or target node features, or the attributes of the edge they are passed along, as indicated by the thickness of incoming arrows.}
    \label{fig:gnn}
\end{figure}

A message passing-type GNN layer is comprised of three functions: 1) a message passing function \textsc{Msg} that permits information exchange between nodes over edges; 2) a permutation-invariant aggregation function \textsc{Agg} that combines the collection of received messages into a single, fixed-length representation; 3) and an update function \textsc{Update} that produces node-level representations given the previous representation and the aggregated messages. Common choices are a simple linear transformation for \textsc{Msg}, summation, simple- or weighted-averages for \textsc{Agg}, and multilayer perceptrons (MLP) with activation functions for the \textsc{Update} function, although it is not uncommon for the \textsc{Msg} or \textsc{Update} function to be absent or reduced to an activation function only. Where the node representations after layer $t$ are $\mathbf{h}^{(t)}$, we have
\begin{align*}
    \textup{\textbf{msg}}_{ji} &= \textsc{Msg}\left( \mathbf{h}^{(t)}_j,\mathbf{h}^{(t)}_i,\mathbf{x}^e_{j,i}\right)\\
    \mathbf{h}^{(t+1)}_i &= \textsc{Update}\left(\mathbf{h}^{(t)}_i,\textsc{Agg}\left(\textup{\textbf{msg}}_{ji},j\in\mathcal{N}_i\right)\right)
\end{align*}
or, more compactly,
\begin{align*}
    \mathbf{h}^{(t+1)}_i &= \gamma^{(t+1)}\left(\mathbf{h}^{(t)}_i,\square_{j\in\mathcal{N}_i}\phi^{(t+1)}\left(\mathbf{h}^{(t)}_i,\mathbf{h}^{(t)}_j,\mathbf{e}_{j,i}\right)\right)
\end{align*}
where $\gamma$, $\square$ and $\phi$ are the update, aggregation and message passing functions, respectively, and $(t)$ indicates the layer index \cite{Fey/Lenssen/2019}. 
\Review{The design of the aggregation $\square$ is important: when chosen to be an injective function, the message passing mechanism can be shown to be equivalent to the color refinement procedure in the Weisfeiler-Lehman algorithm \cite{xu2018powerful}.
}
The initial node representations, $\mathbf{h}^{(0)}_i$, are typically set to node features, $\mathbf{x}^v_i$. Figure \ref{fig:gnn} gives a schematic representation of this operation.

\subsubsection{Graph convolutional network}

The graph convolutional network (GCN) \cite{kipf2016semi} can be decomposed in this framework as
\begin{align*}
    \textsc{Msg}\left(\dots\right) &= \mathbf{W}^{(t)}\mathbf{h}^{(t)}_j \\
    \textsc{Agg}\left(\dots\right) &= \sum_{j \in \mathcal{N}_i} \frac{1}{\sqrt{d_i d_j}}\textup{\textbf{msg}}_{j} \\
    \textsc{Update}\left(\dots\right) &= \sigma\left( \frac{1}{d_i}\mathbf{W}^{(t)}\mathbf{h}^{(t)}_i + \textup{\textbf{agg}}_i\right)
\end{align*}
where $\sigma$ is some activation function, usually a rectified linear unit (ReLU). The scheme is simplified further if we consider the addition of self-loops, that is, an edge from a node to itself, commonly expressed as the modified adjacency $\hat{\mathbf{A}}=\mathbf{A}+\mathbf{I}$, where the aggregation includes the self-message and the update reduces to
\begin{align*}
    \textsc{Update}\left(\dots\right)=\sigma(\textup{\textbf{agg}}_i).
\end{align*}
\noindent With respect to the notations in Figure \ref{fig:gnn}, for GCN we have $\alpha_{i,j}=\frac{1}{\sqrt{d_id_j}}$.

\begin{figure}[ht]
    \centering
    \includegraphics[width=0.9\linewidth]{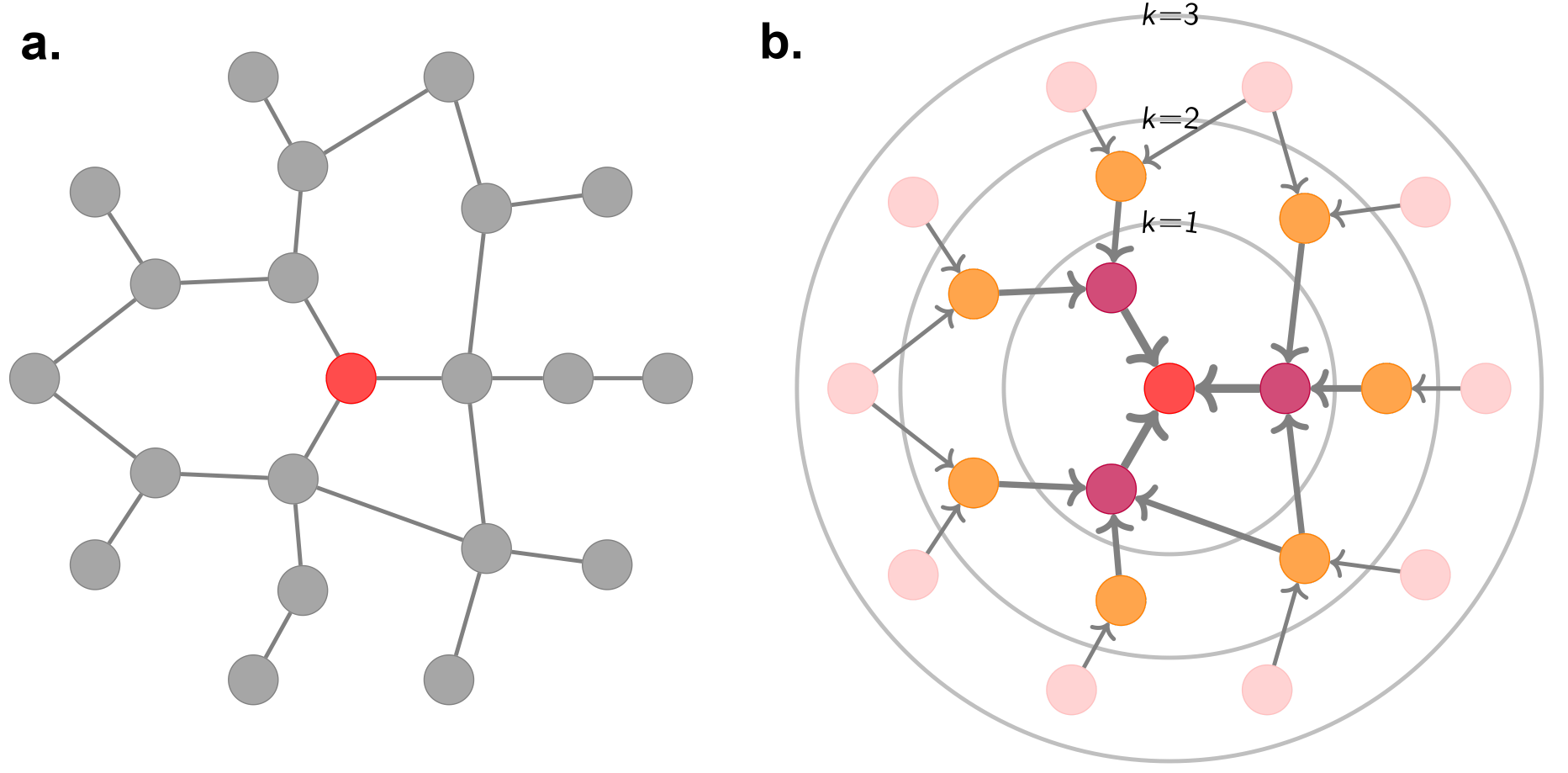}
    \caption{$k$-hop neighbourhoods of the central node (\textcolor{OrangeRed}{red}). Typically, a GNN layer operates on the 1-hop neighbourhood i.e. nodes with which the central node shares an edge, within the $k=1$ circle. Stacking layers allows information from more distant nodes to propagate through intermediate nodes.}
    \label{fig:rec}
\end{figure}

As the update depends only on a node's local neighbourhood, these schemes are also commonly referred to as \textit{neighbourhood aggregation}. Indeed, taking a broader perspective, a single-layer GNN updates a node's features based on its immediate or one-hop neighbourhood. Adding a second GNN layer allows information from the two-hop neighbourhood to propagate via intermediary neighbours. By further stacking GNN layers, node features can come to depend on the initial values of more distant nodes, analogous to the broadening the receptive field in later layers of convolutional neural networks---the deeper the network, the broader the receptive field (see Figure \ref{fig:rec}). However, this process is diffusive and leads to features \textit{washing out} as the graph thermalises. This problem is solved in convolutional networks with pooling layers, but an equivalent canonical coarsening does not exist for irregular graphs.

\subsubsection{Graph attention network}

Graph Attention Networks (GAT) \cite{velivckovic2018graph} weight incoming messages with an attention mechanism and multi-headed attention for train stability. Including self-loops, the message, aggregation and update functions
\begin{align*}
    \textsc{Msg}\left(\dots\right) &= \mathbf{W}^{(t)}\mathbf{h}^{(t)}_j \\
    \textsc{Agg}\left(\dots\right) &= \sum_{j \in \mathcal{N}_i \cup i} \alpha_{ij}\textup{\textbf{msg}}_{j} \\
    \textsc{Update}\left(\dots\right) &= \sigma\left(\textup{\textbf{agg}}_i\right)
\end{align*}
are otherwise unchanged. Although the authors suggest the attention mechanism is decoupled from the architecture and should be task specific, in practice, their original formulation is most widely used. The attention weights, $\alpha_{ij}$ are softmax normalised, that is
\begin{align*}
    \alpha_{ij} = \textup{softmax}_j(e_{ij})= \frac{\exp{(e_{ij})}}{\sum_{k\in\mathcal{N}_i}\exp{(e_{ik})}}
\end{align*}
where $e_{ij}$ is the output of a single layer feed-forward neural network without a bias (a projection) with LeakyReLU activations, that takes the concatenation of transformed source- and target-node features as input,
\begin{align*}
    e_{ij} &= \textsc{MLP}\left([\mathbf{W}^{(t)}\mathbf{h}^{(t)}_i||\mathbf{W}^{(t)}\mathbf{h}^{(t)}_j]\right)\\
    &= \textup{LeakyReLU}\left(\mathbf{a}^\top [\mathbf{W}^{(t)}\mathbf{h}^{(t)}_i||\mathbf{W}^{(t)}\mathbf{h}^{(t)}_j]\right)
\end{align*}
where $\textup{LeakyReLU}(x)=\max(x,\lambda x) ; 0 \leq \lambda \leq 1$.

\subsubsection{Relational graph convolutional networks}

\begin{figure}[ht]
    \centering
    \includegraphics[width=0.9\linewidth]{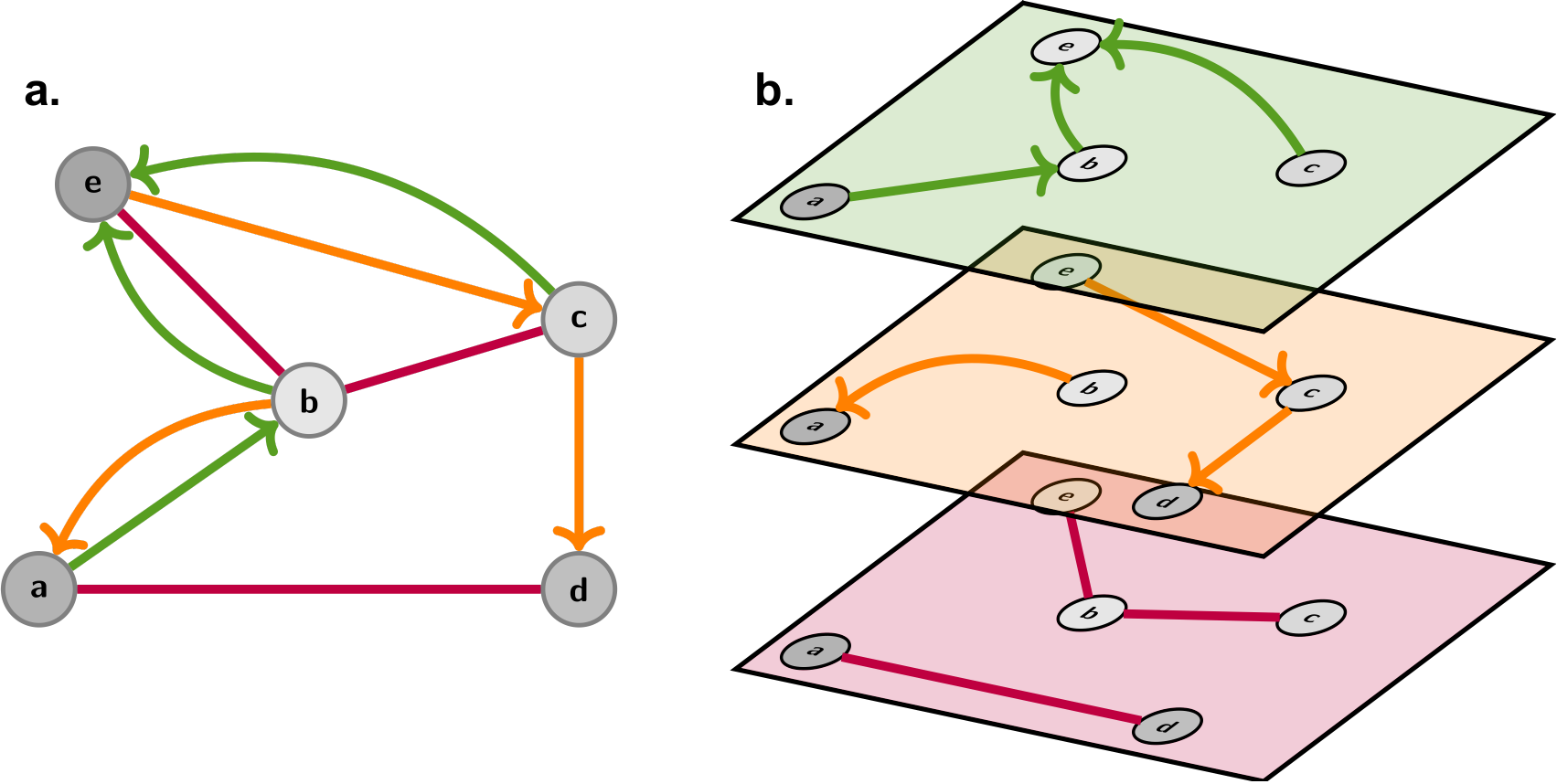}
    \caption{A multi-relational graph \textbf{(a.)} can be parsed into layers of a multiplex graph \textbf{(b.)}. The R-GCN learns a separate transform for each layer, and the self-loop and messages are passed according to the connectivity of the layers. For example, node ($a$) passes a message to node ($b$) in the \textcolor{OliveGreen}{top layer}, receives a message from ($b$) in the \textcolor{orange}{middle layer}, and does not communicate with ($b$) in the \textcolor{RedViolet}{bottom layer}.}
    \label{fig:multiplex}
\end{figure}

At many scales of systems biology, the relationships between entities have a type, a direction, or both. For instance, the type of bonds between atoms, binding of two proteins, and gene regulatory interactions are essential to understanding the systems in which they exist. This idea is expressed in the message passing framework with messages that depend on edge attributes. Relational Graph Convolutional Networks (R-GCNs) \cite{schlichtkrull2018modeling} learn separate linear transforms for each edge type, which can be viewed as casting the graph as a multiplex graph and operating GCN-like models independently on each layer, as shown in Figure \ref{fig:multiplex}.

The R-GCN model decomposes to
\begin{align*}
    \textsc{Msg}_r\left(\dots\right) &= \mathbf{W}^{(t)}_r \mathbf{h}^{(t)}_j \\
    \textsc{Agg}\left(\dots\right) &= \sum_{r \in \mathcal{R}} \sum_{j \in \mathcal{N}^r_i} \frac{1}{c_{i,r}}\textup{\textbf{msg}}^r_{j} \\
    \textsc{Update}\left(\dots\right) &= \sigma\left(\mathbf{W}^{(t)}_0\mathbf{h}^{(t)}_i + \textup{\textbf{agg}}_i\right)
\end{align*}
for edge types $r \in \mathcal{R}$, with separate transforms $\mathbf{W}_0^{(t)}$ for self-loops, and problem-specific normalization constant $c_{i,r}$. 

\Review{The different types of GNNs above illustrate some approaches to define message passing on graphs. Note that there is no established best scheme for all scenarios and that each specific application might require a different scheme. }

\subsubsection{Graph Pooling}

Geometric deep learning approaches machine learning with graph-structured data as the generalisation of methods designed for learning with grid and sequence data (images, time-series; Euclidean data) to non-Euclidean domains, i.e. graphs and manifolds \cite{bronstein2017geometric}. This is also reflected in the derivation and naming conventions of popular graph neural network layers as generalised convolutions \cite{kipf2016semi,schlichtkrull2018modeling}. Modern convolutional neural networks have settled on the combination of layers of $3\times3$ kernels interspersed with $2\times2$ max-pooling. Developing a corresponding pooling workhorse for GNNs is an active area of research. The difficulty is that, unlike Euclidean data structures, there are no canonical up- and down-sampling operations for graphs. As a result, there are many proposed methods that centre around learning to pool or prune based on features \cite{ying2018hierarchical,cangea2018towards,gao2019graph,lee2019self,bodnar2020deep}, and learned or non-parametric structural pooling \cite{boykov2006graph,luzhnica2019clique,bianchi2020mincutpool,jin2020hierarchical}. However, the distinction between featural and structural methods is blurred when topological information is included in the node features.

\begin{figure}[ht]
    \centering
    \includegraphics[width=0.95\linewidth]{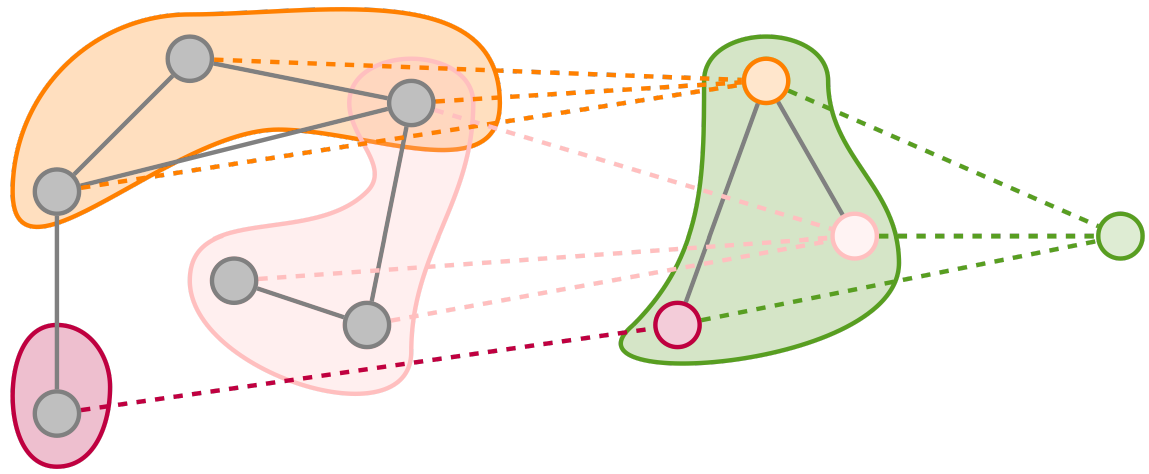}
    \caption{A possible graph pooling schematic. Nodes in the original graph (left, grey) are pooled into nodes in the intermediate graph (centre) as shown by the dotted edges. The final pooling layer aggregates all the intermediate nodes into a single representation (right, green). DiffPool could produce the pooling shown \cite{ying2018hierarchical}.}
    \label{fig:pooling}
\end{figure}

The most successful feature-based methods extract representations from graphs directly either for cluster assignments \cite{ying2018hierarchical,bodnar2020deep} or \texttt{top-k} pruning \cite{cangea2018towards,gao2019graph,lee2019self}. DiffPool uses GNNs both to produce a hierarchy of representations for overall graph classification and to learn intermediate representations for soft cluster assignments to a fixed number of pools \cite{ying2018hierarchical}. Figure \ref{fig:pooling} presents an example of this kind of pooling. \texttt{top-k} pooling takes a similar approach, but instead of using an auxiliary learning process to pool nodes, it is used to prune nodes \cite{cangea2018towards,gao2019graph}. In many settings, this simpler method is competitive with DiffPool at a fraction of the memory cost.

\begin{figure}[ht]
    \centering
    \includegraphics[width=\linewidth]{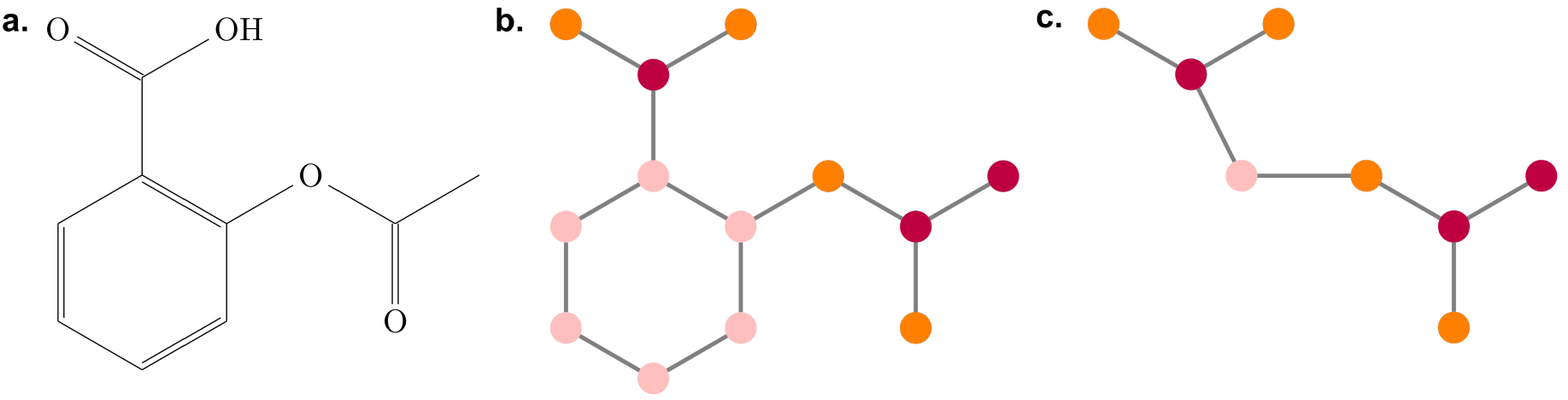}
    \caption{Illustration of \textbf{a.} the molecule aspirin, \textbf{b.} its basic graph representation, and \textbf{c.} the associated junction tree representation. Colours on the node correspond to atom types.}
    \label{fig:aspirin}
\end{figure}

Structure-based pooling methods aggregate nodes based on the graph topology and are often inspired by the processes developed by chemists and biochemists for understanding molecules through their parts. For example, describing a protein in terms of its secondary structure ($\alpha$-helix, $\beta$-sheets) and the connectivity between these elements can be seen as a pooling operation over the protein's molecular graph. Figure \ref{fig:aspirin} shows how a small molecule can be converted to a junction tree representation, with the carbon ring (in pink) being aggregated into a single node. Work on decomposing molecules into motifs bridges the gap between handcrafted secondary structures and unconstrained learning methods \cite{jin2020hierarchical}. Motifs are extracted based on a combined statistical and chemical analysis, where motif templates (i.e. graph substructures) are selected based on how frequently they occur in the training corpus and molecules are then decomposed into motifs according to some chemical rules. More general methods look to concepts from graph theory such as minimum cuts \cite{boykov2006graph,bianchi2020mincutpool} and maximal cliques \cite{luzhnica2019clique} on which to base pooling. Minimum cuts are graph partitions that minimise some objective and have obvious connections to graph clustering, whilst cliques (subsets of nodes that are fully connected) are in some sense at the limit of node community density.

\begin{table*}[h]
\begin{center}
{\small
 \resizebox{.95\textwidth}{!}
 {
 \begin{tabular}{ p{4cm}  p{1.5cm}  p{3cm} p{2cm} p{2cm} p{2cm} p{1.3cm}} 
 \hline
 \textbf{Relevant application} &	\textbf{Reference} &	\textbf{Method type}	 & \textbf{Task level}	& \textbf{ML approach} &	\textbf{Data types}  &	\textbf{Exp. val?} \\
 \hline
 \rowcolor{black!20}
 \ref{s41}\quad Target identification & & & & & &\\
  --- & \cite{pittala2020relation} &	Geometric (\S\ref{s32}) &	Node-level &	Unsupervised & Di, Dr, GA & \\
 \rowcolor{black!20}
 \multicolumn{6}{l}{\ref{s42}\quad Design of small molecules therapies}  &\\
 \multirow{3}{*}{Molecular property prediction} &   \cite{duvenaud2015convolutional} & GNN (\S\ref{s34}) & Graph-level & Supervised & Dr & \\
 & \cite{feinberg2018potentialnet} &	GNN (\S\ref{s34})&	Graph-level	&Supervised& Dr& \\
 & \cite{klicpera2020directional} &	GNN (\S\ref{s34}) &	Graph-level	 & Supervised &  Dr  & \\  \midrule
 Enhanced high throughput screens &	\cite{stokes2020deep} &	GNN (\S\ref{s34}) &	Graph-level &	Supervised & Dr & \checkmark \\  \midrule 
 \multirow{2}{*}{De novo design} & \cite{jin2018junction} &	GNN (\S\ref{s34}) &	Graph-level &	Unsupervised & Dr & \\
 & \cite{zhavoronkov2019deep} &	Factorisation (\S\ref{s33}) &	Graph-level	& Semi-supervised & Dr& \checkmark \\
 \rowcolor{black!20}
 \multicolumn{7}{l}{\ref{s43}\quad Design of new biological entities} \\
 ML-assisted directed evolution	& --- & --- & --- & --- & --- &\\ \midrule
Protein engineering & \cite{gainza2020deciphering} & GNN (\S\ref{s34}) & Subgraph-level$^{*}$ & Supervised & PS & \\ \midrule			
De novo design	& \cite{strokach2020fast} & GNN (\S\ref{s34}) & Graph-level & Supervised & PS & \checkmark \\
 \rowcolor{black!20}
 \ref{s44}\quad Drug repurposing & & & & & &\\
 \multirow{2}{*}{Off-target repurposing} & \cite{olayan2018ddr}  &	 Factorisation (\S\ref{s33})	& Node-level &	Unsupervised & Dr, PI & \\
 & \cite{torng2019graph} &	GNN (\S\ref{s34}) &	Graph-level  &	Supervised  & Dr, PS & \\ \midrule
 \multirow{3}{*}{On-target repurposing}& \cite{yang2019drug} & Factorisation (\S\ref{s33}) &	Node-level &	Unsupervised & Dr, Di & \\
 & \cite{wang2020toward}&	GNN (\S\ref{s34}) &	Node-level  &	Supervised  & Dr, Di & \\ 
 & \cite{zeng2020repurpose} &	Geometric (\S\ref{s32}) &	Node-level &	Unsupervised & Dr, Di, PI, GA & \\ \midrule
 \multirow{2}{*}{Combination repurposing} & \cite{zitnik2018modeling} &	GNN (\S\ref{s34}) &	Node-level  &	Supervised & Dr, PI, DC & \\
 & \cite{jin2020modeling} &	GNN (\S\ref{s34})	& Graph-level 	& Supervised & Dr, DC& \checkmark \\
 \hline
\end{tabular}
}
}
\end{center}
    \caption{Exemplar references linking applications from Section \ref{sec:drugdev} to methods described in Section \ref{mlg}. The entries in the data types column refers to acronyms defined in Table \ref{tab:data}. The last column indicates the presence of follow up experimental validation. $^{*}$The authors consider a mesh graph over protein surfaces.}
    \label{tab:summary}
\end{table*}

\section{Drug development applications}\label{sec:drugdev}

The process of discovering a drug and making it available to patients can take up to 10 years and is characterised by failures, or \textit{attrition}, see Figure \ref{fig:drugdev}. The early discovery stage involves target identification and validation, hit discovery, lead molecule identification and then optimisation to achieve desired characteristics of a drug candidate \cite{hughes2011principles}. Pre-clinical research typically comprises both \textit{in vitro} and \textit{in vivo} assessments on toxicity, pharmacokinetics (PK), pharmacodynamics (PD), and efficacy of the drug. Providing good pre-clinical evidence are presented, the drug then progresses for human clinical trials normally through three different phases of clinical trials. In the following subsections, we explore how GML can be applied to distinct stages within the drug discovery and development process.

\Review{In Table \ref{tab:summary}, we provide a summary of how some of the key work reviewed in this section ties to the methods discussed in Section \ref{mlg}. Table \ref{tab:data} outlines the underlying key data types and databases therein. Biomedical databases are typically presented as general repositories with minimal reference to downstream applications. Therefore, additional processing is often required for a specific task; to this end, efforts have been directed towards processed data repositories with specific endpoints in mind \cite{tdc,biokg}.}

\begin{table}[h]
    \centering
    \begin{tabular}{l l l}
        \hline
        \textbf{Type of data} & \textbf{Databases} & \textbf{Acronym} \\
        \hline
        Drugs {\footnotesize (structure, indications, targets)} & \cite{chembl,drugbank,drughub,pubchem,zinc}  & Dr\\
        Drug combinations & \cite{drugcomb,tatonetti2012data}& DC\\
        Protein {\footnotesize (structure, sequence)} & \cite{pdb}  & PS\\
        Protein interactions & \cite{biogrid,string}  & PI\\
        Gene annotations & \cite{kegg,reactome,go} & GA\\
        Diseases & \cite{do,kegg,reactome,disgenet}  & Di \\
        \hline
    \end{tabular}
    \caption{Different types of data relevant to drug discovery and development applications with associated databases.}
    \label{tab:data}
\end{table}

\subsection{Target identification}\label{sec:target_id}\label{s41}

Target identification is the search for a molecular target with a significant functional role(s) in the pathophysiology of a disease such that a hypothetical drug could modulate said target culminating with beneficial effect \cite{schenone2013target, titov2012identification}. Early targets included G-protein coupled receptors (GPCRs), kinases, and proteases and formed the major target protein families among first-in-class drugs \cite{eder2014discovery} --- targeting other classes of biomolecules is also possible, e.g., nucleic acids. For an organ-specific therapy, an ideal target should be strongly and preferably expressed in the tissue of interest, and preferably a three-dimensional structure should be obtainable for biophysical simulations. 

There is a range of complementary lines of experimental evidence that could support target identification. For example, a phenomenological approach to target identification could consider the imaging, histological, or -omic presentation of diseased tissue when compared to matched healthy samples. Typical differential presentation includes chromosomal aberrations (e.g., from WGS), differential expression (e.g., via RNA-seq) and protein translocation (e.g., from histopathological analysis) \cite{paananen2019omics}. As the availability of -omic technologies increases, computational and statistical advances must be made to integrate and interpret large quantities of high dimensional, high-resolution data on a comparatively small number of samples, occasionally referred to as \textit{panomics} \cite{sandhu2018panomics, matthews2016omics}.

In contrast to a static picture, \textit{in vitro} and \textit{in vivo} models are built to examine the dynamics of disease phenotype to study mechanism. In particular, genes are manipulated in disease models to understand key drivers of a disease phenotype. For example, random mutagenesis could be induced by chemicals or transposons in cell clones or mice to observe the phenotypic effect of perturbing certain cellular pathways at the putative target protein \cite{titov2012identification}. As a targeted approach, bioengineering techniques have been developed to either silence mRNA or remove the gene entirely through genetic editing. In modern times, CRISPR is being used to knockout genes in a \textit{cleaner} manner to prior technologies, e.g. siRNA, shRNA, TALEN \cite{boettcher2015choosing, smith2017evaluation, peretz2018combined}. Furthermore, innovations have led to CRISPR interference (CRISPRi) and CRISPR activation (CRISPRa) that allow for suppression or overexpression of target genes \cite{le2017dual}. 
    
To complete the picture, biochemical experiments observe chemical and protein interactions to inform on possible drug mechanisms of action \cite{schenone2013target}, examples include: affinity chromatography, a range of mass spectrometry techniques for proteomics, and drug affinity responsive target stability (DARTS) assays \cite{cuatrecasas1968selective, lomenick2011target, ong2005mass}. X-ray crystallography and cryogenic electron microscopy (cryo-EM) can be used to detail structures of proteins to identify druggable pockets \cite{shoemaker2018x}; computational approaches can be used to assess the impacts of mutations in cancer resulting in perturbed crystal structures \cite{malhotra2019understanding}. Yeast two-hybrid or three-hybrid systems can be employed to detail genomic protein--protein or RNA--protein interactions \cite{hamdi2012yeast, licitra1996three}. 

Systems biology aims to unify phenomenological observations on disease biology (the ``-omics view''), genetic drivers of phenotypes (driven by bioengineering) through a network view of interacting biomolecules \cite{butcher2004systems}. The ultimate goal is to pin down a ``druggable'' point of intervention that could hopefully reverse the disease condition. One of the outcomes of this endeavour is the construction of signalling pathways; for example, the characterisation of the TGF-$\beta$, PI3K/AKT and Wnt-dependent signalling pathways have had profound impacts on oncology drug discovery \cite{akhurst2012targeting,hennessy2005exploiting,janssens2006wnt}.

In contrast to complex diseases, target identification for infectious disease requires a different philosophy. After eliminating pathogenic targets structurally similar to those within the human proteome, one aims to assess the druggability of the remaining targets. This may be achieved using knowledge of the genome to model the constituent proteins when 3D structures are not already available experimentally. The Blundell group has shown that 70-80\% of the proteins from \textit{Mycobacterium tuberculosis} and a related organism, \textit{Mycobacterium abscessus} (infecting cystic fibrosis patients), can be modelled via homology \cite{ochoa2015chopin, skwark2019mabellini}. By examining the potential binding sites, such as the catalytic site of an enzyme or an allosteric regulatory site, the binding hotspot can be identified and the potential value as a target estimated \cite{blundell2020personal}. Of course, target identification is also dependent on the accessibility of the target to a drug, as well as the presence of efflux pumps --- and metabolism of any potential drug by the infectious agent.

\subsubsection{From systems biology to machine learning on graphs}


Organisms, or biological systems, consist of complex and dynamic interactions between entities at multiple scales. At the submolecular level, proteins are chains of amino acid residues which fold to adopt highly specific conformational structures. At the molecular scale, proteins and other biomolecules physically interact through transient and long-timescale binding events to carry out regulatory processes and perform signal amplification through cascades of chemical reactions. By starting with a low-resolution understanding of these biomolecular interactions, canonical sequences of interactions associated with specific processes become labelled as \textit{pathways} that ultimately control cellular functions and phenotypes. Within multicellular organisms, cells interact with each other forming diverse tissues and organs. A reductionist perspective of disease is to view it as being the result of perturbations of the cellular machinery at the molecular scale that manifest through aberrant phenotypes at the cellular and organismal scales. Within target identification, one is aiming to find nodes that upon manipulation lead to a causal sequence of events resulting in the reversion from a diseased to a healthy state.

It seems plausible that target identification will be the greatest area of opportunity for machine learning on graphs. From a genetics perspective, examining Mendelian traits and genome-wide association studies (GWAS) linked to coding variants of drug targets have a greater chance of success in the clinic \cite{king2019drug, nelson2015support}. However, when examining protein--protein interaction networks, Fang \textit{et al.} \cite{fang2019genetics} found that various protein targets were not themselves ``genetically associated'', but interacted with other proteins with genetic associations to the disease in question. For example in the case of rheumatoid arthritis (RA), tumour necrosis factor (TNF) inhibition is a popular drug mechanism of action with no genetic association --- but the interacting proteins of TNF including CD40, NFKBIA, REL, BIRC3, and CD8A are known to exhibit a genetic predisposition to RA.

Oftentimes, systems biology has focused on networks with static nodes and edges, ignoring faithful characterisation of underlying biomolecules that the nodes represent. With GML, we can account for much richer representations of biology accounting for multiple relevant scales, for example, graphical representation of molecular structures (discussed in Sections \ref{sec:design_NCE} and \ref{sec:design_NBE}), functional relationships within a knowledge graph (discussed in Section \ref{sec:repurposing}), and expression of biomolecules. Furthermore, GML can learn graphs from data as opposed to relying on pre-existing incomplete knowledge \cite{wang2019dynamic,kazi2020differentiable}. Early work utilising GML for target identification includes Pittala \textit{et al.} \cite{pittala2020relation}, whereby a knowledge graph link prediction approach was used to beat the in house algorithms of Open Targets \cite{carvalho2019open} to rediscover drug targets within clinical trials for Parkinson's Disease.

The utilisation of multi-omic expression data capturing instantaneous multimodal snapshots of cellular states will play a significant role in target identification as costs decrease \cite{nicora2020integrated, sanchez2020interpreting} --- particularly in a precision medicine framework \cite{sandhu2018panomics}. Currently, however, only a few panomic datasets are publicly available. A small number of early adopters have spotted the clear utility in employing GML \cite{wang2020moronet}, occasionally in a multimodal learning \cite{nguyen2020multiview, ma2019integrate}, or causal inference setting \cite{pfister2019stabilizing}. These approaches have helped us move away from the classical Mendelian ``one gene -- one disease'' philosophy and appreciate the true complexity of biological systems.

\subsection{Design of small molecule therapies}\label{sec:design_NCE}\label{s42}

Drug design broadly falls into two categories: phenotypic drug discovery and target-based drug discovery. Phenotypic drug discovery (PDD) begins with a disease's phenotype without having to know the drug target. Without the bias from having a known target, PDD has yielded many first-in-class drugs with novel mechanisms of action \cite{swinney2011were}. It has been suggested that PDD could provide the new paradigm of drug design, saving a substantial amount of costs and increasing the productivity \cite{moffat2017opportunities}. However, drugs found by PDD are often pleiotropic and impose greater safety risks when compared to target-oriented drugs. In contrast, best-in-class drugs are usually discovered by a target-based approach.

For target-based drug discovery, after target identification and validation, ``hit'' molecules would be identified via high-throughput screening of compound libraries against the target \cite{hughes2011principles}, typically resulting in a large number of possible hits. Grouping these into ``hit series'' and they become further refined in functional \textit{in vitro} assays. Ultimately, only those selected via secondary \textit{in vitro} assays and \textit{in vivo} models would be the drug ``leads''. With each layer of screening and assays, the remaining compounds should be more potent and selective against the therapeutic target. Finally, lead compounds are optimised by structural modifications, to improve properties such as PKPD, typically using heuristics, e.g. Lipinski's rule of five \cite{lipinski1997experimental}. In addition to such structure-based approach, fragment-based (FBDD) \cite{blundell2002high, murray2010structural} and ligand-based drug discovery (LBDD) have also been popular \cite{erlanson2004fragment,acharya2011recent}. FBDD enhances the ligand efficiency and binding affinity with fragment-like leads of approximately $\sim$150 Da, whilst LBDD does not require 3D structures of the therapeutic targets.

Both phenotypic and target-based drug discovery comes with their own risks and merits. While the operational costs of target ID may be optional, developing suitable phenotypic screening assays for the disease could be more time-consuming and costly \cite{zheng2013phenotypic,moffat2017opportunities}. Hence, the overall timeline and capital costs are roughly the same \cite{zheng2013phenotypic}. 
 
In this review, we make no distinction between new chemical entities (NCE), new molecular entities (NME), or new active substances (NAS) \cite{branch2014new}. 

\subsubsection{Modelling philosophy}

For a drug, the base graph representation is obtained from the molecule's SMILES signature and captures bonds between atoms, i.e. each node of the graph corresponds to an atom and each edge stands for a bond between two atoms \cite{duvenaud2015convolutional,klicpera2020directional}. The features associated with atoms typically include its element, valence, and degree. Edge features include the associated bond's type (single, double, triple), its aromaticity, and whether it is part of a ring or not. Additionally, Klicpera \textit{et al.} \cite{klicpera2020directional} consider the geometric length of a bond and geometric angles between bonds as additional features. This representation is used in most applications, sometimes complemented or augmented with heuristic approaches.

To model a graph structure, Jin \textit{et al.} \cite{jin2018junction} used the base graph representation in combination with a \textit{junction tree} derived from it. To construct the junction tree, the authors first define a set of molecule substructures, such as rings. The graph is then decomposed into overlapping components, each corresponding to a specific substructure. Finally, the junction tree is defined with each node corresponding to an identified component and each edge associates overlapping components. 

Jin \textit{et al.} then extended their previous work by using a hierarchical representation with various coarseness of the small molecule \cite{jin2020hierarchical}. The proposed representation has three levels: 1) an atom layer, 2) an attachment layer, and 3) a motif layer. The first level is simply the basic graph representation. The following levels provide the coarse and fine-grain connection patterns between a molecule's motifs. Specifically, the attachment layer describes at which atoms two motifs connect, while the motif layer only captures if two motifs are linked. Considering a molecule base graph representation $G=(\mathcal{V},\mathcal{E})$, a motif is defined as a subgraph of $G$ induced on atoms in $\mathcal{V}$ and bonds in $\mathcal{E}$. Motifs are extracted from a molecule's graph by breaking \textit{bridge bonds}. 

Kajino \cite{kajino2019molecular} opted for a hypergraph representation of small molecules. A hypergraph is a generalisation of graphs in which an edge, called a hyperedge, can connect any number of nodes. In this setting, a node of the hypergraph corresponds to a bond between two atoms of the small molecule. In contrast, a hyperedge then represents an atom and connects all its bonds (i.e. nodes).

\subsubsection{Molecular property prediction} 

Pharmaceutical companies may screen millions of small molecules against a specific target, e.g., see GlaxoSmithKline's DNA-encoded small molecule library of 800 million entries \cite{clark2009design}. However, as the end result will be optimised via a skilled medicinal chemist, one should aim to substantially cut down the search space by screening only a representative selection of molecules for optimisation later. One route towards this is to select molecules with heterogeneous chemical properties using GML approaches. This is a popular task with well-established benchmarks such as QM9 \cite{ramakrishnan2014quantum} and MD17 \cite{chmiela2017machine}. Top-performing methods are based on graph neural networks.

For instance, using a graph representation of drugs, Duvenaud \textit{et al.} \cite{duvenaud2015convolutional} have shown substantial improvements over non-graph-based approaches for molecule property prediction tasks. Specifically, the authors used GNNs to embed each drug and tested the predictive capabilities of the model on diverse benchmark datasets. They demonstrated improved interpretability of the model and predictive superiority over previous approaches which relied on circular fingerprints \cite{glen2006circular}. The authors use a simple GNN layer with a read-out function on the output of each GNN layer that updates the global drug embedding. 

Alternatively, Schutt \textit{et al.} \cite{schutt2018schnet} introduced SchNet, a model that characterises molecules based on their representation as a list of atoms with interatomic distances, that can be viewed as a fully connected graph. SchNet uses learned embeddings for each atom using two modules: 1) an atom-wise module, and 2) an interaction module. The former applies a simple MLP transformation to each atom representation input, while the latter updates the atom representation based on the representations of the other atoms of the molecule and using relative distances to modulate contributions. The final molecule representation is obtained with a global sum pooling layer over all atoms' embeddings.

With the same objective in mind, Klicpera \textit{et al.} \cite{klicpera2020directional} recently introduced DimeNet, a novel GNN architecture that diverges from the standard message passing framework presented in Section \ref{mlg}. DimeNet defines a message coefficient between atoms based on their relative positioning in 3D space. Specifically, the message from node $v_j$ to node $v_i$ is iteratively updated based on $v_j$'s incoming messages as well as the distances between atoms and the angles between atomic bonds. DimeNet relies on more geometric features, considering both the angles between different bonds and the distance between atoms. The authors report substantial improvements over SOTA models for the prediction of molecule properties on two benchmark datasets.

Most relevant to the later stages of preclinical work, Feinberg \textit{et al.} extended previous work on molecular property prediction \cite{feinberg2018potentialnet} to include ADME properties \cite{feinberg2020improvement}. In this scenario, by only using structures of drugs predictions were made across a diverse range of experimental observables, including half-lives across \textit{in vivo} models (rat, dog), human Ether-\`{a}-go-go-Related Gene (hERG) protein interactions and IC$_{50}$ values for common liver enzymes predictive of drug toxicity.

\subsubsection{Enhanced high throughput screens} 

Within the previous section, chemical properties were \textit{a priori} defined. In contrast, Stokes \textit{et al.} \cite{stokes2020deep} leveraged results from a small phenotypic growth inhibition assay of 2,335 molecules against \textit{Escherichia coli} to infer antibiotic properties of the ZINC15 collection of $>$107 million molecules. After ranking and curating hits, only 23 compounds were experimentally tested --- leading to \textit{halicin} being identified. Of particular note was that the Tanimoto similarity of halicin when compared its nearest neighbour antibiotic, metronidazole, was only $\sim$0.21 --- demonstrating the ability of the underlying ML to generalise to diverse structures.

Testing halicin against a range of bacterial infections, including \textit{Mycobacterium tuberculosis}, demonstrated broad-spectrum activity through selective dissipation of the $\Delta$pH component of the proton motive force. In a world first, Stokes \textit{et al.}  showed efficacy of an AI-identified molecule \textit{in vivo} (\textit{Acinetobacter baumannii} infected neutropenic BALB/c mice) to beat the standard of care treatment (metronidazole) \cite{stokes2020deep}. 

\subsubsection{\textit{De novo} design} 

A more challenging task than those previously discussed is \textit{de novo} design of small molecules from scratch; that is, for a fixed target (typically represented via 3D structure) can one design a suitable and \textit{selective} drug-like entity? 

In the landmark paper, Zhavoronkov \textit{et al.} \cite{zhavoronkov2019deep} created a novel chemical matter against discoidin domain receptor 1 (DDR1) using a variational autoencoder style architecture. Notably, they penalise the model to select structures similar to disclosed drugs from the patent literature. As the approach was designed to find small molecules for a well-known target, crystal structures were available and subsequently utilised. Additionally, the ZINC dataset containing hundreds of millions of structures was used (unlabelled data) along with confirmed positive and negative hits for DDR1. 

In total, 6 compounds were synthesised with 4 attaining $<$1$\mu\text{M}$ $\text{IC}_{50}$ values. Whilst selectivity was shown for 2 molecules of DDR1 when compared to DDR2, selectivity against a larger panel of off-targets was not shown. Whilst further development (e.g., PK or toxicology testing) was not shown, Zhavoronkov \textit{et al.} demonstrated \textit{de novo} small molecule design in an experimental setting \cite{zhavoronkov2019deep}. Arguably, the recommendation of an existing molecule is a simpler task than designing one from scratch.

\subsection{Design of new biological entities}\label{sec:design_NBE}\label{s43}

New biological entities (NBE) refer to biological products or biologics, that are produced in living systems \cite{shire2009formulation}. The types of biologics are very diversified, from proteins ($>$40 amino acids), peptides, antibodies, to cell and gene therapies. Therapeutic proteins tend to be large, complex structured and are unstable in contrast to small molecules \cite{patel2015biologics}. Biologic therapies typically use cell-based production systems that are prone to post-translational modification and are thus sensitive to environmental conditions requiring mass spectrometry to characterise the resulting heterogeneous collection of molecules \cite{mo2012structural}.

In general, the target-to-hit-to-lead pathway also applies to NBE discovery, with similar procedures like high-throughput screening assays. Typically, an affinity-based high-throughput screening method is used to select from a large library of candidates using one target. One must then separately study off-target binding from similar proteins, peptides and immune surveillance \cite{kumar2020characterization}. 

\subsubsection{Modelling philosophy}

\begin{figure}
    \centering
    \includegraphics[width=0.95\linewidth]{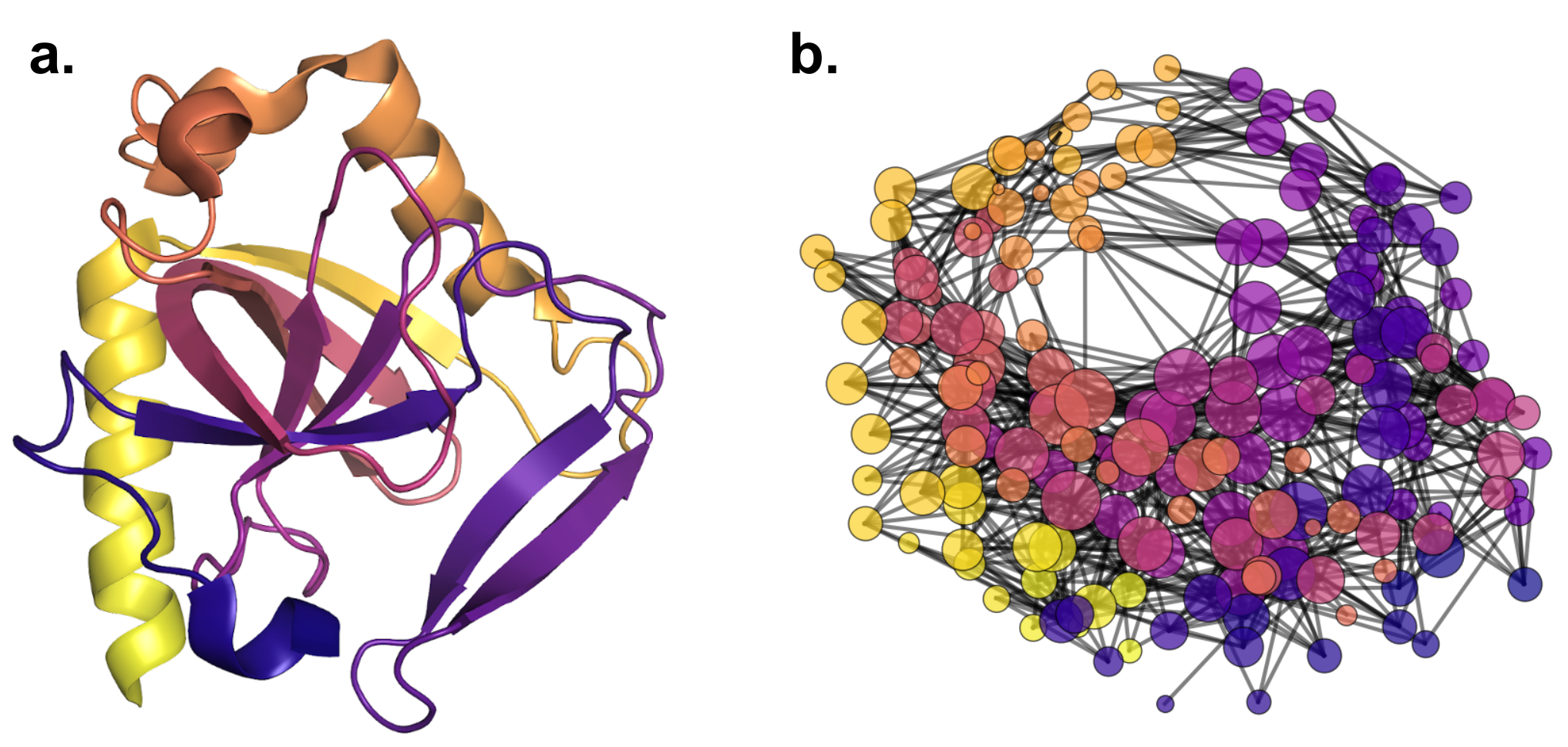}
    \caption{Illustration of \textbf{a.} a protein (PDB accession: 3EIY) and \textbf{b.} its graph representation derived based on intramolecular distance with cut-off threshold set at $10$\AA.}
    \label{fig:proteinasgraph}
\end{figure}

Focusing on proteins, the consensus to derive the protein graph representation is to use pairwise spatial distances between amino acid residues, i.e. the protein's \textit{contact map}, and to apply an arbitrary cut-off or Gaussian filter to derive adjacency matrices \cite{fout2017protein,zamora2019structural,gligorijevic2020structure}, see Figure \ref{fig:proteinasgraph}. 

However, protein structures are substantially more complex than small molecules and, as such, there are several resulting graph construction schemes. For instance, residue-level graphs can be constructed by representing the intramolecular interactions, such as hydrogen bonds, that compose the structure as edges joining their respective residues. This representation has the advantage of explicitly encoding the internal chemistry of the biomolecule, which determines structural aspects such as dynamics and conformational rearrangements. Other edge constructions can be distance-based, such as K-NN (where a node is joined to its $k$ most proximal neighbours) \cite{fout2017protein, zamora2019structural} or based on Delaunay triangulation. Node features can include structural descriptors, such as solvent accessibility metrics, encoding the secondary structure, distance from the centre or surface of the structure and low-dimensional embeddings of physicochemical properties of amino acids. It is also possible to represent proteins as large molecular graphs at the atomic level in a similar manner to small molecules. Due to the plethora of graph construction and featurisation schemes available, tools are being made available to facilitate the pre-processing of said protein structure \cite{Jamasb2020}. 

One should note that sequences can be considered as special cases of graphs and are compatible with graph-based methods. However, in practice, language models are preferred to derive protein embeddings from amino acids sequences \cite{rao2019evaluating,rives2019biological}. Recent works suggest that combining the two can increase the information content of the learnt representations \cite{ingraham2019generative,gligorijevic2020structure}. Several recurrent challenges in the scientific community aim to push the limit of current methods. For instance, the CAFA \cite{radivojac2013large} and CAPRI \cite{lensink_capri2020} challenges aim to improve protein functional classification and protein--protein interaction prediction.

\subsubsection{ML-assisted directed evolution} 

Display technologies have driven the modern development of NBEs; in particular, phage display and yeast display are widely used for the generation of therapeutic antibodies. In general, a peptide or protein library with diverse sequence variety is generated by PCR, or other recombination techniques \cite{galan2016library}. The library is ``displayed'' for genotype-phenotype linkage such that the protein is expressed and fused to surface proteins while the encoding gene is still encapsulated within the phage or cell. Therefore, the library could be screened and selected, in a process coined ``biopanning'', against the target (e.g. antigen) according to binding affinity. Thereafter, the selected peptides are further optimised by repeating the process with a refined library. In phage display, selection works by physical capture and elution \cite{nixon2014drugs}; for cell-based display technologies (like yeast display), fluorescence-activated cell sorting (FACS) is utilised for selection \cite{bradbury2011beyond}.

Due to the repeated iteration of experimental screens and the high number of outputs, such display technologies are now being coupled to ML systems for greater speed, affinity and further \textit{in silico} selection \cite{rickerby2020machine, yang2019machine}. As of yet, it does not appear that advanced GML architectures have been applied in this domain, but promising routes forward have been recently developed. For example, Hawkins-Hooker \textit{et al.} \cite{hawkins2020generating} trained multiple variational autoencoders on the amino acid sequences of 70,000 luciferase-like oxidoreductases to generate new functional variants of the \textit{luxA} bacterial luciferase. Testing these experimentally led to variants with increased enzyme solubility without disrupting function. Using this philosophy, one has a grounded approach to refine libraries for a directed evolution screen.

\subsubsection{Protein engineering}

Some proteins have reliable scaffolds that one can build upon, for example, antibodies whereby one could modify the variable region but leave the constant region intact. For example, Deac \textit{et al.} \cite{deac2019attentive} used dilated (\'{a} trous) convolutions and self-attention on antibody sequences to predict the paratope (the residues on the antibody that interact with the antigen) as well as a cross-modal attention mechanism between the antibody and antigen sequences. Crucially, the attention mechanisms also provide a degree of interpretability to the model.

In a protein-agnostic context, Fout \textit{et al.} \cite{fout2017protein} used Siamese architectures \cite{bromley1994signature} based on GNNs  to predict at which amino acid residues are involved in the interface of a protein--protein complex. Each protein is represented as a graph where nodes correspond to amino acid residues and edges connect each residue to its $k$ closest residues. The authors propose multiple aggregation functions with varying complexities and following the general principles of a Diffusion Convolutional Neural Network \cite{atwood2016diffusion}. The output embeddings of each residue of both proteins are concatenated all-to-all and the objective is to predict if two residues are in contact in the protein complex based on their concatenated embeddings. The authors report a significant improvement over the method without the GNN layers, i.e. directly using the amino acid residue sequence and structural properties (e.g. solvent accessibility, distance from the surface).

Gainza \textit{et al.} \cite{gainza2020deciphering} recently introduced Molecular Surface Interaction Fingerprinting (MaSIF) for tasks such as protein--ligand prediction, protein pocket classification, or protein interface site prediction. The approach is based on GML applied on mesh representations of the solvent-excluded protein surface, abstracting the underlying sequence and internal chemistry of the biomolecule. In practice, MaSIF first discretises a protein surface with a mesh where each point (vertex) is considered as a node in the graph representation. Then the protein surface is decomposed into overlapping small patches based on the geodesic radius, i.e. clusters of the graph. For each patch, geometric and chemical features are handcrafted for all nodes within the patch. The patches serve as bases for learnable Gaussian kernels \cite{monti2017geometric} that locally average node-wise patch features and produce an embedding for each patch. The resulting embeddings are fed to task-dependent decoders that, for instance, give patch-wise scores indicating if a patch overlaps with an actual protein binding site.

\subsubsection{\textit{De novo} design}

One of the great ambitions of bioengineering is to design proteins from scratch. In this case, one may have an approximate structure in mind, e.g., to inhibit the function of another endogenous biomolecule. This motivates the inverse protein folding problem, identifying a sequence that can produce a pre-determined protein structure. For instance, Ingraham \textit{et al.} \cite{ingraham2019generative} leveraged an autoregressive self-attention model using graph-based representations of structures to predict corresponding sequences. 

Strokach \textit{et al.} \cite{strokach2020fast} leveraged a deep graph neural network to tackle protein design as a constraint satisfaction problem. Predicted structures resulting from novel sequences were initially assessed \textit{in silico} using molecular dynamics and energy-based scoring. Subsequent \textit{in vitro} synthesis of sequences led to structures that matched the secondary structure composition of serum albumin evaluated using circular dichroism.

With a novel amino acid sequence that could generate the desired shape of an arbitrary protein, one would then want to identify potential wanted and unwanted effects via functional annotations. These include enzyme commission (EC) numbers, a hierarchy of labels to characterise the reactions that an enzyme catalyses --- previously studied by the ML community \cite{ryu2019deep, dalkiran2018ecpred}.

Zamora \textit{et al.} \cite{zamora2019structural} developed a pipeline based on graph representations of a protein's amino acid residues for structure classification. The model consists of the sequential application of graph convolutional blocks. A block takes as input two matrices corresponding to residue features and coordinates, respectively. The block first uses a layer to learn Gaussian filters applied on the proteins' spatial distance kernel, hence deriving multiple graph adjacency matrices. These are then used as input graphs in a GNN layer which operates on the residue features. The block then performs a 1D average pooling operation on both the feature matrix and the input coordinate matrix, yielding the outputs of the block. After the last block, the final feature matrix is fed to a global attention pooling layer which computes the final embedding of the protein used as input to an MLP for classification. The model performs on par with existing 3D-CNN-based models. However, the authors observe that it is more interpretable, enabling the identification of substructures that are characteristic of structural classification.

Recently, Gligorijevic \textit{et al.} \cite{gligorijevic2020structure} proposed DeepFRI, a model that predicts a protein's functional annotations based on structure and sequence information. The authors define the graph representation of a protein-based on the contact map between its residues. They first use a pre-trained language module that derives $\mathbf{X}_V$, each protein's graph is then fed to multiple GCN layers \cite{kipf2016semi}. The outputs of each GCN layer are concatenated to give the final embedding of a protein that is fed to an MLP layer giving the final functional predictions. The authors report substantial improvements over SOTA methods. Furthermore, they highlight the interpretability of their approach, demonstrating the ability to associate specific annotations with particular structures.

\subsection{Drug repurposing}\label{sec:repurposing}\label{s44}

The term drug repurposing is used to describe the use of an existing drug, whether approved or in development as a therapy, for an indication other than the originally intended indication. Considering that only $12\%$ of drugs that reach clinical trials receive FDA approval, repurposed drugs offer an attractive alternative to new therapies as they are likely to have shorter development times and higher success rates with early stages of drug discovery already completed. It has been estimated that repurposed treatments account for approximately 30\% of newly FDA approved drugs and their associated revenues \cite{pillaiyar2020medicinal}, and that up to 75\% of entities could be repurposed \cite{nosengo2016new}. Note that we incorporate product line extensions (PLEs) within drug repurposing whereby one wishes to identify secondary indications, different formulations for an entity, or partner drugs for combination therapies.

As such, there is a major interest in using \textit{in silico} methods to screen and infer new treatment hypotheses \cite{hodos2016silico}. Drug repurposing relies on finding new indications for existing molecules, either by identifying actionable pleiotropic activity (off-target repurposing), similarities between diseases (on-target repurposing), or by identifying synergistic combinations of therapies (combination repurposing). Well-known examples include: ketoconazole (Nizoral) used to treat fungal infections via enzyme CYP51A1 and now used to treat Cushing syndrome via off-target interactions with CYP17A1 (steroid synthesis/degradation), NR3C4 (androgen receptor), and NR3C1 (glucocorticoid receptor); sildenafil (Viagra) originally intended for pulmonary arterial hypertension on-target repurposed to treat erectile dysfunction; and Genvoya, a combination of emtricitabine, tenofovir alafenamide (both reverse transcriptase inhibitors), elvitegravir (an integrase inhibitor), and cobicistat (a CYP3A inhibitor to improve PK) to treat human immunodeficiency virus (HIV).

\subsubsection{Off-target repurposing}

Estimates suggest that each small molecule may interact with tens, if not hundreds of proteins \cite{zhou2015comprehensive}. Due to small molecule pleiotropy --- particularly from first-in-class drugs \cite{moffat2017opportunities} --- off-targets of existing drugs can be a segue to finding new indications.

A variety of traditional techniques are used to identify missing drug--target interactions. For instance, when the structure of the protein is known, these stem from biophysical simulations, i.e. molecular docking or molecular dynamics. Depending on how one thresholds sequence similarity between the $\sim$21,000 human proteins, structural coverage whereby a 3D structure of the protein exists ranges from 30\% ($\geq98\%$ seq. sim.) to 70\% ($\geq30\%$ seq. sim.) \cite{somody2017structural}. However, $\sim$34\% of proteins are classified as intrinsically disordered proteins (IDPs) with no 3D structure \cite{deiana2019intrinsically, uversky2019intrinsically}. Besides, drugs seldom have a fixed shape due to the presence of rotatable bonds. There is now a growing body of GML literature to infer missing drug--target interactions both with and without relying on the availability of a 3D protein structure.

Requiring protein structures, Torng \textit{et al.} \cite{torng2019graph} focused on the task of associating drugs with protein pockets they can bind to. Drugs are represented based on their atomic structures and protein pockets are characterised with a set of key amino acid residues connected based on Euclidean distance. Drug embeddings are obtained with the GNN operator from Duvenaud \textit{et al.} \cite{duvenaud2015convolutional}. To derive embeddings of protein pockets, the authors first use two successive graph autoencoders with the purposes of 1) deriving a compact feature vector for each residue, and 2) deriving a graph-level representation of the protein pocket itself. These autoencoders are pre-trained, with the encoder of the first serving as input to the second. Both encoders are then used as input layers of the final model. The association prediction between a drug and a protein pocket is then obtained by feeding the concatenation of the drug and pocket representations to an MLP layer. The authors report improved performance against the previous SOTA model based on a 3D-CNN operating on a grid-structure representation of the protein pocket \cite{ragoza2017protein}.

A range of GML methods for drug--target interaction do not require protein structure. For instance, Gao \textit{et al.} \cite{gao2018interpretable} use two encoders to derive embeddings for proteins and drugs, respectively. For the first encoder, recurrent neural networks are used to derive an embedding matrix of the protein-based on its sequence and functional annotations. For the second encoder, each drug is represented by its underlying graph of atoms and the authors use GNNs to extract an embedding matrix of the graph. They use three layers of a Graph Isomorphism Network \cite{xu2018powerful} to build their subsequent architecture. Finally, a global attention pooling mechanism is used to extract vector embeddings for both drugs and proteins based on their matrix embeddings. The two resulting vectors are fed into a Siamese neural network \cite{bromley1994signature} to predict their association score. The proposed approach is especially successful compared to baseline for cold-start problems where the protein and/or drug are not present in the training set.

Alternatively, Nascimento \textit{et al.} \cite{nascimento2016multiple} introduce KronRLS-MKL, a method that casts drug--target interaction prediction as a link prediction task on a bi-partite graph capturing drug--protein binding. The authors define multiple kernels capturing either drug similarities or protein similarities based on multiple sources of data. The optimisation problem is posed as a multiple kernel learning problem. Specifically, the authors use the Kronecker operator to obtain a kernel between drug--protein pairs. The kernel is then used to predict a drug--protein association based on their similarity to existing drug--target link. Crichton \textit{et al.} \cite{crichton2018neural} cast the task in the same setting. However, the authors use existing embedding methods, including node2vec \cite{grover2016node2vec}, deepwalk \cite{perozzi2014deepwalk}, and LINE \cite{tang2015line}, to embed nodes in a low-dimensional space such that the embeddings capture the local graph topology. The authors feed these embeddings to a machine learning model trained to predict interactions. The underlying assumption is that a drug will be embedded closer to its protein targets.

Similarly, Olayan \textit{et al.} \cite{olayan2018ddr} propose DDR to predict drug--target interactions. The authors first build a graph where each node represents either a drug or a protein. In addition to drug--protein edges, an edge between two drugs (or two proteins) represents their similarity according to a predefined heuristic from multiple data sources. DDR embeds each drug--protein pair based on the number of paths of predefined types that connect them within the graph. The resulting embeddings are fed to a random forest algorithm for drug--target prediction. 

\Review{Recently, Mohamed \textit{et al.} \cite{mohamed2020discovering} proposed an end-to-end knowledge graph embedding model to identify off-target interactions. The authors construct a large knowledge graph encompassing diverse data pertaining to drugs and proteins, such as associated pathways and diseases. Using an approach derived from DistMult and ComplEx, the authors report state-of-the-art results for off-target prediction.}

\subsubsection{On-target repurposing}

On-target repurposing takes a holistic perspective and uses known targets of a drug to infer new putative indications based on diverse data. For instance, one can identify functional relationships between a drug's targets and genes associated with a disease. Also, one may look for similarities between diseases --- especially those occurring in different tissues. Hypothetically, one could prospectively find repurposing candidates in the manner of Fang \textit{et al.} \cite{fang2019genetics} by finding a missing protein--protein interactions between a genetically validated target and a drug's primary target. Knowledge graph completion approaches have been particularly effective in addressing these tasks.

For instance, Yang \textit{et al.} \cite{yang2019drug} introduced Bounded Nuclear Norm Regularisation (BNNR). The authors build a block matrix with a drug similarity matrix, a disease similarity matrix, and a disease--drug indication matrix. The method is based on the matrix completion property of Singular Value Thresholding algorithm applied to the block matrix. BNNR incorporates regularisation terms to balance approximation error and matrix rank properties to handle noisy drug–drug and disease–disease similarities. It also adds a constraint that clips the association scores to the interval $[0,1]$. The authors report performance improvements when compared to competing approaches.

Alternatively, Wang \textit{et al.} \cite{wang2020toward} recently proposed an approach to predict new drug indications based on two bipartite graphs, capturing drug--target interactions and disease--gene associations, and a PPI graph. Their algorithm is composed of an encoder module, relying on GAT \cite{velivckovic2018graph}, and an MLP decoder module. The encoder derives drug and disease embeddings through the distillation of information along the edges of the graphs. The input features for drugs and diseases are based on similarity measures. On the one hand, drug features correspond to the Tanimoto similarities between its SMILES representation and that of the other drugs. On the other hand, a disease's features are defined by its similarity to other diseases computed based on MeSH-associated terms.

\subsubsection{Combination repurposing}

Combination drugs have been particularly effective in diseases with complex aetiology or an evolutionary component where resistance to treatment is common, such as infectious diseases. If synergistic drugs are found, one can reduce dose whilst improving efficacy \cite{keith2005multicomponent, he2018methods}. Strategically, combination therapies provide an additional way to extend the indications and efficacy of available entities. They can be used for a range of purposes, for example, convenience and compliance as a fixed-dose formulation (e.g. valsartan and hydrochlorothiazide for hypertension \cite{dipette2019fixed}), to achieve synergies (e.g. co-trimoxazole: trimethoprim and sulfamethoxazole for bacterial infections), to broaden spectrum (e.g. for treatment of infections by an unknown pathogen), or to combat disease resistance (e.g. multi-drug regimens for drug-sensitive and drug-resistant tuberculosis). The number of potential pairwise combinations of just two drugs makes a brute force empirical laboratory testing approach a lengthy and daunting prospect. To give a rough number, there exist around 4,000 approved drugs which would require $\sim$8 million experiments to test all possible combinations of two drugs at a single dose. Besides, there are limitless ways to change the dosage and the timing of treatments, as well as the delivery method.

Arguably some of the first work using GML to model combination therapy was DECAGON by Zitnik \textit{et al.} \cite{zitnik2018modeling} used to model polypharmacy side-effects via a multi-modal graph capturing drug--side effect--drug triplets in addition to PPI interactions. In contrast, Deac \textit{et al.} \cite{deac2019drug} forwent incorporation of a knowledge graph instead modelling drug structures directly and using a coattention mechanism to achieve a similar level of accuracy. Typically architectures predicting drug--drug antagonism can be minimally adapted for prediction of synergy. However, more nuanced architectures are emerging combining partial knowledge of drug--target interactions with target--disease machine learning modules \cite{jin2020modeling}.

\subsubsection{Outlook}

In the last year to address the unprecedented COVID-19 global health crisis, multiple research groups have explored graph-based approaches to identify drugs that could be repurposed to treat SARS-CoV-2 \cite{zhou2020network,morselli2020network,zeng2020repurpose,ioannidis2020few}. For instance, Morselli \textit{et al.} \cite{morselli2020network} proposed an ensemble approach combining three different graph-based association strategies. The first two are similar in principle. First, each drug and each disease is represented by the set of proteins that it targets. Second, the association between a drug and a disease is quantified based on the distance between the two sets on a PPI graph. The two approaches differ on whether the distance is quantified with shortest paths or random walks. The last strategies rely on GNNs for multimodal graphs (knowledge graph). The graph contains PPIs, drug--target interactions, disease--protein associations, and drug indications. The formulation of the GNN layer is taken from the DECAGON model \cite{zitnik2018modeling}, an architecture similar to the R-GCN model \cite{schlichtkrull2018modeling}.

Alternatively, Zeng \textit{et al.} \cite{zeng2020repurpose} use RotatE to identify repurposing hypotheses to treat SARS-CoV-2 from a large knowledge graph constructed from multiple data sources and capturing diverse relationships between entities such as drugs, diseases, genes, and anatomies. Additionally, Ioannidis \textit{et al.} \cite{ioannidis2020few} proposed a modification of the RGCN architecture to handle few-shot learning settings in which some relations only connect a handful of nodes.

In the rare disease arena, Sosa \textit{et al.} \cite{sosa2019literature} introduced a knowledge graph embedding approach to identify repurposing hypotheses for rare diseases. The problem is cast as a link prediction problem in a knowledge graph. The authors use the Global Network of Biological Relationships (GNBR) \cite{percha2018global}, a knowledge graph built through literature mining and that contains diverse relationships between diseases, drugs, and genes. Due to the uncertainty associated with associations obtained from literature mining, the authors use a knowledge graph embedding approach design to account for uncertainty \cite{chen2019embedding}. Finally, the highest-ranking associations between drugs and rare diseases are investigated, highlighting literature and biological support.

Drug repurposing is now demonstrating itself as a first use case of GML methods likely to lead to new therapies within the coming years. Outside of the pharmaceutical industry, GML methods recommending nutraceuticals \cite{veselkov2019hyperfoods} may also offer fast routes to market through generally recognized as safe (GRAS) regulatory designations.

\section{Discussion}\label{sec:discussion}

We have discussed how GML has produced the state-of-the-art results both on graph-level problems for the description of drugs and other biomolecules, and node-level problems for the navigation of knowledge graphs and representation of disease biology. With the design, synthesis and testing of \textit{de novo} small molecules \cite{zhavoronkov2019deep}, the \textit{in vitro} and \textit{in vivo} testing of drug repurposing hypotheses \cite{stokes2020deep}, and target identification frameworks being conceptualised \cite{pittala2020relation}, we are potentially entering into a golden age of validation for GML within drug discovery and development.

A few key hurdles limit lossless representation of biology. At the molecular level, bonds can be either rotatable single bonds or fixed bonds; accurately representing the degrees of freedom of a molecule is a topic of active research \cite{flam2020neural}. At the cellular level, expression of mRNA and proteins exhibit stochastic dynamics \cite{kholodenko2006cell, raj2008nature}. A pathway is not expressed in a binary fashion: some proteins may only have the potential to be expressed, e.g. via unspliced pre-mRNA, and meanwhile, proteases are actively recycling unwanted proteins. Historically, most -omic platforms have recorded an average ``bulk'' signal; however, with the recent single-cell revolution, GML offers a principled approach to the characterisation of signalling cascades.

GML is still in its infancy, and underlying theoretical guarantees and limitations are under active research. For instance, deeper GNNs suffer from oversmoothing of features and oversquashing of information. Oversmoothing is the phenomenon of features \textit{washing out} through repeated rounds of message passing and aggregation \cite{oono2019graph}. The inverse, having too few layers to exchange information globally, is referred to as under-reaching \cite{barcelo2019logical}. These issues have limited the expressivity of traditional GNNs \cite{dehmamy2019understanding,chen2020can}. To alleviate this, a promising direction is to incorporate global information in the model, for instance, by using contrastive approaches \cite{velickovic2019deep,sun2019infograph}, by augmenting the framework with relative positioning to anchor nodes \cite{you2019position,li2020distance}, or by implementing long-range information flow \cite{flam2020neural}.

Due to the problem of missing data within biomedical knowledge graphs, we envisage opportunities for \textit{active learning} to be deployed to label critical missing data points to explore experimentally and therefore reduce model uncertainty \cite{sverchkov2017review}. Due to the significant expense associated with drug discovery and development, integrating \textit{in silico} modelling and experimental research is of great strategic importance. While active learning has previously led to biased datasets in other settings, modern techniques are addressing these drawbacks \cite{gudovskiy2020deep, aggarwal2020active}.

Finally, because GML allows for the representation of unstructured multimodal datasets, one can expect to see tremendous advances made within data integration. Most notably, highly multiplexed single-cell omic technologies are now being expanded in spatial settings \cite{burgess2019spatial, baharlou2019mass}. In addition, CRISPR screening data with associated RNA sequencing readouts are emerging as promising tools to identify key genes controlling a cellular phenotype \cite{daniloski2020identification}.

\section{Acknowledgements}

We gratefully acknowledge support from William L. Hamilton, Benjamin Swerner, Lyuba V. Bozhilova, and Andrew Anighoro.

\bibliographystyle{unsrt}
\bibliography{references}
\end{document}